\documentclass[linenumbers]{aastex631}
\usepackage{CJK}
\usepackage{amsmath}
\usepackage{bm}
\usepackage{multirow,amsmath,array,booktabs}
\usepackage{color}
\usepackage{ulem}

\newcommand{\ra}{\rangle}

\newcommand{\bbm}{\begin{bmatrix}}
\newcommand{\ebm}{\end{bmatrix}}
\newcommand{\bBm}{\begin{Bmatrix}}
\newcommand{\eBm}{\end{Bmatrix}}
\newcommand{\bpm}{\begin{pmatrix}}
\newcommand{\epm}{\end{pmatrix}}

\newcommand{\nc}{\newcommand}       
\nc{\vc}[1] {\mbox{\boldmath $#1$}} 
\nc{\del}       {\partial}              
\nc{\bra}       {\langle}               
\nc{\ket}       {\rangle}               
\nc{\bras}[1]   {\langle #1|}           
\nc{\kets}[1]   {|#1\rangle}            
\nc{\mapleft}[1]{           
	\smash{\mathop{\,          %
			\hbox to 1.5cm{\rightarrowfill}\, }\limits_{#1}}}
\nc{\beq}     {\begin{eqnarray}} \nc{\eeq}    {\end{eqnarray}}
\nc{\nn}      {\\\nonumber} \nc{\vs}      {\vspace{-0.275cm}}
\nc{\fra}    {\frac{1}{2}}
\nc{\mb}        {\mathbf}

\graphicspath{{./}{figures/}}

\begin{document}
\begin{CJK*}{UTF8}{gbsn}

\title{Nuclear Matter and Neutron Stars from Relativistic Brueckner-Hartree-Fock Theory}

\author{Hui Tong (童辉)}
\affiliation{College of Physics and Materials Science, Tianjin Normal University, Tianjin 300387, China}
\affiliation{Strangeness Nuclear Physics Laboratory, RIKEN Nishina Center, Wako, 351-0198, Japan}

\author{Chencan Wang (王宸璨)}
\affiliation{School of Physics, Nankai University, Tianjin 300071, China}

\author{Sibo Wang (王锶博)}
\email{sbwang@cqu.edu.cn}
\affiliation{Department of Physics, Chongqing University, Chongqing 401331, China}

\begin{abstract}
  The momentum and isospin dependence of the single-particle potential for the
  in-medium nucleon are the key quantities in the Relativistic Brueckner-Hartree-Fock (RBHF) theory.
  It depends on how to extract the scalar and the vector components of the
  single-particle potential inside nuclear matter.
  In contrast to the RBHF calculations in the Dirac space with the positive-energy
  states (PESs) only, the single-particle potential can be determined in a unique way
  by the RBHF theory together with the negative-energy states (NESs),
  i.e., the RBHF theory in the full Dirac space. The saturation properties of symmetric and asymmetric nuclear matter in the full Dirac space
   are systematically investigated based on the realistic Bonn nucleon-nucleon potentials.
  In order to further specify the importance of the calculations in the
  full Dirac space, the neutron star properties are investigated.
  The direct URCA process in neutron star cooling will happen at density
  $\rho_{\rm{DURCA}}=0.43,~0.48,~0.52$~fm$^{-3}$ with the proton fractions $Y_{p,\rm{DURCA}}=0.13$.
  The radii of a $1.4M_\odot$ neutron star are predicated as $R_{1.4M_\odot}=11.97,~12.13,~12.27$~km, and
  their tidal deformabilities are $\Lambda_{1.4M_\odot}=376,~405,~433$ for potential Bonn A,~B,~C.
  Comparing with the results obtained in the Dirac space with PESs only, full-Dirac-space RBHF calculation predicts the softest symmetry energy which would be
  more favored by the gravitational waves (GW) detection from GW170817.
  Furthermore, the results from full-Dirac-space RBHF theory are consistent with the recent
  astronomical observations of massive neutron stars and simultaneous mass-radius measurement.
\end{abstract}



\section{Introduction}\label{Introduction}

Compact matter in the universe, such as neutron stars, provides an intriguing interplay
between nuclear processes and astrophysical observables~\citep{Lattimer2004,BURGIO2021PPNP}.
The central density in massive neutron stars reaches several times the empirical nuclear matter saturation density $(\rho_0 \approx 0.16~\textrm{fm}^{-3})$, which is far from
those encountered in terrestrial experiments.
Thus, astrophysical observations of neutron star properties are crucial in constraining the equation
of state (EOS) for dense nuclear matter, since the latter constitutes the basic input quantity for
the theoretical reconstruction of a neutron star.

During the past decade, considerable progress has been achieved with precise measurements
for the massive neutron stars PSR~J1614-2230~($1.928\pm0.017M_\odot$)~\citep{demorest2010,Fonseca2016APJ},
PSR~J0348+0432~($2.01\pm0.04M_\odot$)~\citep{antoniadis2013}, and PSR~J0740+6620~($2.14_{-0.09}^{+0.10}
M_\odot$)~\citep{cromartie2020}.
Recently, for the first time the mass and radius of PSR~J0030+0451 were simultaneously measured by the
Neutron star Interior Composition Explorer (NICER) collaboration ~\citep{Raaijmakers_2019}.
From NICER's data, two independent analysis groups have reported a mass of $1.44_{-0.14}^{+0.15}M_\odot$
with a radius of $13.02_{-1.06}^{+1.24}$km~\citep{Miller2019} and a mass of $1.34_{-0.16}^{+0.15}M_\odot$
with a radius of $12.71_{-1.19}^{+1.14}$km~\citep{Riley_2019}.
In addition, the advanced LIGO and Virgo Collaborations detected a gravitational wave (GW) signal from a
binary neutron star (BNS) merger, i.e., GW170817~\citep{Abbott2017PRL}.
This has greatly advanced multimessenger astronomy and stimulated an intense research activity in the
field of nuclear physics.
Besides the masses and radii of BNS, a significant signature carried by GW is the tidal deformability
which represents the mass quadrupole moment response of a neutron star to the strong gravitational field
induced by its companion~\citep{Damour1992,Hinderer2008,Flanagan2008,Damour2009,Postnikov2010}.
In particular, the limits on the tidal deformability inferred from the GW signal have been widely used to
constrain the neuron star radius~\citep{Fattoyev2018PRL,Annala2018,Most2018,Tews2018}, the neutron skin
thickness of $^{208}$Pb~\citep{Fattoyev2018PRL}, and the nuclear matter
EOS~\citep{Malik2018,NaiBoZhang2018APJ,Tong2020PRC}.
Two years after GW170817, the detection of a compact binary coalescence involving a $22.2-24.3 M_\odot$
black hole and a compact object with a mass of $2.50-2.67 M_\odot$ was reported by the LIGO and Virgo
Collaborations, i.e., GW190814~\citep{Abbott2020APJ}.
The secondary object of GW190814 still challenges the physics community nowadays, since it may be either
the heaviest neutron star or the lightest black hole ever
discovered~\citep{FattoyevPRC2020,VattisPRD2020,Safarzadeh2020ApJ,Yang2020ApJ,Tsokaros2020ApJ,HuangApJ2020,
BombaciPRL2021}.

Theoretically, many attempts have also been made to obtain the EOS of neutron star matter via different nuclear
many-body theories.
They can roughly be divided into two categories: density functional theories (DFTs) with
effective nucleon-nucleon $(NN)$ interactions and \textit{ab-initio} methods with realistic ones.

The effective $NN$ interactions in DFTs, either non-relativistic or  relativistic, are constructed to
reproduce the properties of finite nuclei and nuclear matter around saturation density, such as
Skyrme~\citep{Skyrme1956,Vautherin1972_PRC5-626}, Gogny~\citep{Decharge1980_PRC21-1568}, or relativistic
mean-field (RMF) models~\citep{Boguta1977_NPA292-413,
Reinhard1989_RPP52-439,Ring1996_PPNP37-193,Meng2006_PPNP57-470,CDFT2016}. However, since the isovector
channels of the density functionals are loosely constrained in the fitting procedures, their predictions for
the nuclear matter properties such as symmetry energy at higher densities still have large
uncertainties~\citep{LiChenKo2008,Jiang2015PRC}.

In \textit{ab-initio} methods, the realistic $NN$ interactions are adopted, such as Bonn~\citep{Machleidt1989} potentials determined by fitting the $NN$ scattering data.
The relativistic Bonn potential has been successfully applied in relativistic
Brueckner-Hartree-Fock (RBHF) theory~\citep{Shen2019}, to study both nuclear matter and neutron
stars~\citep[e.g.][]{HOROWITZ1984,Brockmann1990,Engvik1994,Jong1998,Alonso2003,VanDalen2005_PRL95-022302,Krastev2006,Katayama2013,Tong2018,Tong2020PRC}. Comparing with non-relativistic BHF theory, it is well known
that the RBHF theory can reproduce the empirical saturation properties of symmetric nuclear matter (SNM)
without an extra three-body force~\citep{Brockmann1990}.

In the RBHF calculations, one of the most important procedure is the self-consistent
determination of the single-particle potential of the nucleons with the effective $G$ matrix inside the nuclear medium.
According to the limitations of symmetries~\citep{SerotWalecka1986}, the single-particle
potential operator $\mathcal{U}$ is generally divided into scalar and vector components.
However, due to the complexity of the procedure, there exists different approaches to extract the components of
single-particle potential in the RBHF theory, i.e., the momentum-independence
approximation~\citep{Brockmann1990}, the projection
method~\citep{HOROWITZ1987,NUPPENAU1989,GROSSBOELTING1999} and the RBHF theory in the full Dirac
space~\citep{WANG-SB2021_PRC103-054319}.

In the momentum-independence approximation~\citep{Brockmann1990}, the momentum dependence of the scalar potential $U_{S}$ and the timelike component of vector potential $U_{0}$ are neglected by regarding them as constants at a given baryon density.
When investigating the isospin asymmetric nuclear matter (ANM), this method can not determine the correct behavior of the isospin dependence of the single-particle potential~\citep{Ulrych1997_PRC56-1788,Schiller2001}.
In the projection method~\citep{HOROWITZ1987,NUPPENAU1989,GROSSBOELTING1999}, the effective $G$ matrix is projected to Lorentz invariant amplitudes which are used to determine the scalar and vector components of the single-particle potential.
This method is more accurate than the momentum-independence approximation, since it keeps the spacelike component of the vector potential and the momentum dependence of the single-particle potential.
However, this method is still restricted to the positive-energy states (PESs) without considering the negative-energy states (NESs).
In paticular, contradictory results for the isospin dependence of the single-particle potential are found between the two approximations~\citep{Ulrych1997_PRC56-1788}.

Recently, to avoid the approximations used in the momentum-independence approximation and the projection method in the Dirac space with PESs only, a new self-consistent RBHF calculation in the full Dirac space for both SNM and ANM has been developed~\citep{WANG-SB2021_PRC103-054319,SiboWang2022arxiv}.
The Lorentz structure and the momentum dependence of the single-particle potential are determined uniquely by decomposing the matrix elements of $\mathcal{U}$ in the full Dirac space.
In particular, the long-standing controversy of the isospin dependence of the single-particle potential has been clarified for the opposite tendency predicted by the momentum-independence approximation and projection method.

In this work, it is timely and necessary to study the nuclear matter properties above the saturation density and neutron star properties with the latest RBHF theory in the full Dirac space using the Bonn potentials.
In order to compare with the results obtained in the Dirac space with PESs only, we also perform the calculations with the momentum-independence approximation and the projection method.

This paper is arranged as follows. The theoretical
framework of the RBHF theory in the full Dirac space, the nuclear matter properties, and the neutron star properties are reviewed in Section~\ref{Theoreticalframework}. In Section~\ref{Resultsdiscussions}, the properties of nuclear matter and neutron star are presented and discussed. The summary is given in Section~\ref{summary}.

\section{Theoretical framework}\label{Theoreticalframework}

\subsection{Relativistic Brueckner-Hartree-Fock (RBHF) theory in the full Dirac space}

In the RBHF theory, the essential point is to describe the single-particle motion in nuclear matter by using the Dirac equation
\begin{equation}\label{DiracEquation}
  \left\{ \bm{\alpha}\cdot\bm{p}+\beta \left[M+\mathcal{U}_\tau(\bm{p})\right] \right\} u_\tau(\bm{p},s)
  = E_{\bm{p},\tau}u_{\tau}(\bm{p},s), \quad \tau = n,p,
\end{equation}
where $\bm{\alpha}$ and $\beta$ are the Dirac matrices, $M$ is the nucleon mass, $\bm{p}$ and $E_{\bm{p},\tau}$ are the momentum and single-particle energy, $s$ and $\tau$ denote the spin and isospin.
According to the translational and rotational invariance, time-reversal invariance, hermiticity, and parity conservation, the single-particle potential $\mathcal{U}_\tau$ in the infinite, uniform nuclear matter can be expressed as~\citep{SerotWalecka1986}
\begin{equation}\label{SPP}
  \mathcal{U}_\tau(\bm{p}) = U_{S,\tau}(p)+ \gamma^0U_{0,\tau}(p) + \bm{\gamma\cdot\hat{p}}U_{V,\tau}(p),
\end{equation}
where $U_{S,\tau}$ is the scalar part of single-particle potential, $U_{0,\tau}$ and $U_{V,\tau}$ are the timelike and spacelike parts of the vector potential.
$\hat{\bm{p}}=\bm{p}/|\bm{p}|$ is the unit vector.
It should be emphasized that the single-particle potential in Eq.~\eqref{SPP} is the function of the nucleon momentum $\bm{p}$.

With the definitions of the following effective quantities:
\begin{subequations}
	\begin{align}
    	\bm{p}^*_\tau=&\ \bm{p}+\hat{\bm{p}}U_{V,\tau}(p),\label{eq:eff-p}\\
    	M^*_{\bm{p},\tau}=&\ M+U_{S,\tau}(p),\label{eq:eff-m}\\
    	E^*_{\bm{p},\tau}=&\ E_{\bm{p},\tau}-U_{0,\tau}(p)\label{eq:eff-e},
    \end{align}
\end{subequations}
the solutions of Eq.~\eqref{DiracEquation} in the full Dirac space are
\begin{equation}\label{DiracSpinor}
  u_\tau(\bm{p},s) =\ \sqrt{\frac{E_{\bm{p},\tau}^*+M_{\bm{p},\tau}^*}{2M_{\bm{p},\tau}^*}}
  		\bbm 1 \\ \frac{\bm{\sigma}\cdot\bm{p}^*_\tau}{E_{\bm{p},\tau}^*+M_{\bm{p},\tau}^*}\ebm \chi_s\chi_\tau,~~~v_\tau(\bm{p},s) =\ \gamma^5u_\tau(\bm{p},s),
\end{equation}
where $u_\tau$ and $v_\tau$ are in-medium baryon spinors with positive and negative energy, $\chi_s$ and $\chi_\tau$ are the spin and isospin wave functions.
The baryon spinor can be obtained once the single-particle potentials are determined.
The single-particle potentials in Eq.~\eqref{SPP} can be obtained uniquely from three matrix elements $\Sigma^{++}_\tau$, $\Sigma^{-+}_\tau$, and $\Sigma^{--}_\tau$  of the operator $\mathcal{U}_\tau$ in the full Dirac space~\citep{Anastasio1981_PRC23-2273,Poschenrieder1988_PRC38-471},
\begin{subequations}\label{Sigma2US0V}
  \begin{align}
    U_{S,\tau}(p) = &\ \frac{\Sigma^{++}_\tau(p)-\Sigma^{--}_\tau(p)}{2},\\
    U_{0,\tau}(p) = &\ \frac{E^*_{\bm{p},\tau}}{M^*_{\bm{p},\tau}}\frac{\Sigma^{++}_\tau(p)+\Sigma^{--}_\tau(p)}{2}
    					 - \frac{p^*_\tau}{M^*_{\bm{p},\tau}}\Sigma^{-+}_\tau(p),\\
    U_{V,\tau}(p) = &\ -\frac{p^*_\tau}{M^*_{\bm{p},\tau}}\frac{\Sigma^{++}_\tau(p)+\Sigma^{--}_\tau(p)}{2}
    				  + \frac{E^{*}_{\bm{p},\tau}}{M^*_{\bm{p},\tau}} \Sigma^{-+}_\tau(p).
  \end{align}
\end{subequations}
This indicates that one needs the matrix elements not only for the positive-energy solutions given in Eq.~\eqref{DiracSpinor}, but also the elements coupling positive- with negative-energy solutions given in Eq.~\eqref{DiracSpinor} and those for the negative-energy solutions, i.e., $\Sigma^{++}_\tau$, $\Sigma^{-+}_\tau$, and $\Sigma^{--}_\tau$.
These three matrix elements can be evaluated in the the mean-field approximation through the effective $NN$ interactions $G$ matrix in the full Dirac space
\begin{subequations}\label{Gm2Sigma}
  \begin{align}
    \Sigma^{++}_\tau(p) = &\ \sum_{s'\tau'} \int^{k^{\tau'}_F}_0 \frac{d^3p'}{(2\pi)^3}
    			\frac{M^*_{\bm{p}',\tau'}}{E^*_{\bm{p}',\tau'}}
    			\langle \bar{u}_\tau(\bm{p},1/2) \bar{u}_{\tau'}(\bm{p}',s')| \bar{G}^{++++}(W)|
    			u_\tau(\bm{p},1/2)u_{\tau'}(\bm{p}',s')\ra, \label{Sigma++} \\
    \Sigma^{-+}_\tau(p) = &\ \sum_{s'\tau'} \int^{k^{\tau'}_F}_0 \frac{d^3p'}{(2\pi)^3}
    			\frac{M^*_{\bm{p}',\tau'}}{E^*_{\bm{p}',\tau'}}
			    \langle \bar{v}_\tau(\bm{p},1/2) \bar{u}_{\tau'}(\bm{p}',s')| \bar{G}^{-+++}(W)|
			    u_\tau(\bm{p},1/2)u_{\tau'}(\bm{p}',s')\ra, \label{Sigma-+} \\
    \Sigma^{--}_\tau(p) = &\ \sum_{s'\tau'} \int^{k^{\tau'}_F}_0 \frac{d^3p'}{(2\pi)^3}
    			\frac{M^*_{\bm{p}',\tau'}}{E^*_{\bm{p}',\tau'}}
    			\langle \bar{v}_\tau(\bm{p},1/2) \bar{u}_{\tau'}(\bm{p}',s')| \bar{G}^{-+-+}(W)|
    			v_\tau(\bm{p},1/2)u_{\tau'}(\bm{p}',s')\ra, \label{Sigma--}
  \end{align}
\end{subequations}
where $k_F^{\tau'}$ specifies the Fermi momentum for neutrons and protons, which has a relation to the total baryonic density $\rho=\rho_n + \rho_p$.
$\bar{G}$ is the antisymmetrized $G$ matrix, where $\pm$ in the superscript denotes PESs or NESs.

To derive the effective interactive $G$ matrix in nuclear matter, one of the most widely used equations in the RBHF theory is the in-medium covariant Thompson equation~\citep{Brockmann1990},
\begin{equation}\label{ThomEqu}
	\begin{split}
  G_{\tau\tau'}(\bm{q}',\bm{q}|\bm{P},W)
  =&\ V_{\tau\tau'}(\bm{q}',\bm{q}|\bm{P})
  + \int \frac{d^3k}{(2\pi)^3}
  V_{\tau\tau'}(\bm{q}',\bm{k}|\bm{P}) \\
    & \times \frac{M^{*}_{\bm{P}+\bm{k},\tau}M^{*}_{\bm{P}-\bm{k},\tau'}}{E^*_{\bm{P}+\bm{k},\tau}E^*_{ \bm{P}-\bm{k},\tau'}}
    \frac{Q_{\tau\tau'}(\bm{k},\bm{P})}{W-E_{\bm{P}+\bm{k},\tau}-E_{\bm{P}-\bm{k},\tau'}}  G_{\tau\tau'}(\bm{k},\bm{q}|\bm{P},W),
  \end{split}
\end{equation}
where $\tau\tau'=nn,~pp$ or $np$, $W$ is the starting energy, $V_{\tau\tau'}$ denotes a realistic bare $NN$ interaction and it is constructed in terms of effective Dirac spinors in the full Dirac space as explained in Eq.~\eqref{DiracSpinor}.
Three parametrizations of Bonn potentials, namely, Bonn A, B, and C~\citep{Machleidt1989} are chosen here.
$\bm{P}=\frac{1}{2}({\bm k}_1+{\bm k}_2)$ is the center-of-mass momentum and $\bm{k}=\frac{1}{2}({\bm k}_1-{\bm k}_2)$ is the relative momentum of the two interacting nucleons with momenta ${\bm k}_1$ and ${\bm k}_2$.
$\bm{q}, \bm{k}$ and $\bm{q}'$ are the initial, intermediate, and final relative momenta of the two nucleons scattering in nuclear matter, respectively. $Q_{\tau\tau'}(\bm{k},\bm{P})$ is the Pauli operator, which prevents $NN$ scattering into occupied states in the nuclear medium.


The relativistic $G$ matrix is self-consistently calculated with the single-particle potentials in the standard RBHF iterative procedure through equations~\eqref{DiracEquation}, \eqref{Sigma2US0V}, \eqref{Gm2Sigma}, and \eqref{ThomEqu}.
Once the iteration has converged, the binding energy per nucleon in nuclear matter can be calculated by
\begin{equation}\label{E/A}
  \begin{split}
  E/A
  =&\ \frac{1}{\rho} \sum_{s,\tau} \int^{k^\tau_F}_0 \frac{d^3p}{(2\pi)^3} \frac{M^*_{\bm{p},\tau}}{E^*_{\bm{p},\tau}}
  \langle \bar{u}_\tau(\bm{p},s)| \bm{\gamma}\cdot\bm{p} + M |u_\tau(\bm{p},s)\ra - M \\
    &\ + \frac{1}{2\rho} \sum_{s,s',\tau,\tau'} \int^{k^\tau_F}_0 \frac{d^3p}{(2\pi)^3} \int^{k^{\tau'}_F}_0
     	  \frac{d^3p'}{(2\pi)^3} \frac{M^*_{\bm{p},\tau}}{E^*_{\bm{p},\tau}}\frac{M^*_{\bm{p}',\tau'}}{E^*_{\bm{p}',\tau'}} \\
  	&\ \times \langle \bar{u}_\tau(\bm{p},s) \bar{u}_{\tau'}(\bm{p}',s') |\bar{G}^{++++}(W)| u_\tau(\bm{p},s) u_{\tau'}(\bm{p}',s') \ra.
  \end{split}
\end{equation}

\subsection{Nuclear matter properties}

The binding energy per nucleon of ANM can be generally expressed as a power series in the asymmetry parameter $\alpha=(\rho_{n}-\rho_{p})/\rho$,
\begin{equation}\label{binding_energy}
  E/A(\rho,\alpha)=E/A(\rho,0)+E_{\mathrm{sym}}(\rho)\alpha^{2}+\mathcal{O}(4),
\end{equation}
where $E/A(\rho,0)$ is the binding energy per nucleon in SNM and $E_{\mathrm{sym}}(\rho)$ is the nuclear symmetry energy,
\begin{equation}
  E_{\mathrm{sym}}(\rho)=\frac{1}{2}\frac{\partial^{2}E/A(\rho,\alpha)}%
  {\partial\alpha^{2}}\bigg|_{\alpha=0}.
  \label{equ2}
\end{equation}

The binding energy per nucleon in SNM can be expanded around the saturation density $\rho_{0}$,
\begin{equation}
  E/A(\rho,0)=E/A(\rho_{0},0)+\frac{K_{\infty}}{2}\left(  \frac{\rho-\rho_{0}}%
  {3\rho_{0}}\right)  ^{2}+\mathcal{O}(3),
  \label{equ3}
\end{equation}
where $E/A(\rho_{0},0)$ denotes the binding energy per nucleon at $\rho_0$.
The second derivative of $E/A(\rho,0)$ with respect to $\rho$ is given by the incompressibility $K(\rho)$,
\begin{equation}\label{equ4}
  K(\rho)=9\rho^{2}\frac{\partial^{2}E/A(\rho,0)}{\partial\rho^{2}},
\end{equation}
and $K_{\infty}$ is the value at $\rho_{0}$.
Around the saturation density $\rho_{0}$, the symmetry energy can be similarly expanded as
\begin{equation}
  E_{\mathrm{sym}}(\rho)=E_{\mathrm{sym}}(\rho_{0})+L\left(  \frac{\rho-\rho
  _{0}}{3\rho_{0}}\right)  + \frac{K_{\mathrm{sym}}}{2}\left(  \frac{\rho
  -\rho_{0}}{3\rho_{0}}\right)  ^{2}+\mathcal{O}(3),
  \label{equ8}
\end{equation}
where $E_{\mathrm{sym}}(\rho_{0})$ is the value of the symmetry energy at saturation density, $L$ and $K_{\mathrm{sym}}$ are the slope parameter and curvature parameter of the nuclear symmetry energy at $\rho_{0}$:
\begin{eqnarray}%
  L&=&3\rho\frac{\partial E_{\mathrm{sym}}(\rho)}{\partial\rho}%
  \bigg|_{\rho=\rho_{0}},
  \label{equ9}\\
  K_{\mathrm{sym}}&=&9\rho^{2}\frac{\partial^{2} E_{\mathrm{sym}}(\rho
  )}{\partial\rho^{2}}\bigg|_{\rho=\rho_{0}}.
  \label{equ10}
\end{eqnarray}
Furthermore, $K_{\mathrm{asy}}=K_{\mathrm{sym}}-6L$ has been widely used to describe the isospin dependence of the incompressibility of ANM in
Refs.~\citep{BARAN2005,ChenKoLi2005,LiChenKo2008,Sun2008,Centelles2009}.


\subsection{Neutron star properties}

The neutron star is described as the $\beta$-stable
nuclear matter system in this work, which consists of nucleons and leptons (mainly $e^-$ and $\mu^-$).
The chemical potentials of nucleons and leptons satisfy the equilibrium conditions
\begin{equation}\label{chemequilb}
  \mu_p = \mu_n - \mu_e, \quad \mu_\mu=\mu_e,
\end{equation}
where $\mu_e$, $\mu_\mu$, $\mu_p$, and $\mu_n$ are the chemical potentials of leptons, proton and neutron.
In addition, charge neutrality is required as
\begin{equation}\label{chargneut}
  \rho_p = \rho_e + \rho_\mu.
\end{equation}
With these constraints, the energy density of the $\beta$-stable nuclear matter is then obtained as
\begin{equation}\label{nm-ene}
  \varepsilon = \rho \left[E(\rho,1-2Y_p)/A + Y_p M_p + (1-Y_p)M_n \right]+\varepsilon_e+\varepsilon_\mu,
\end{equation}
where the leptons are treated as non-interacting Fermi gas~\citep{Bombaci1991,Maieron2004} and $Y_i = \rho_i/\rho~(i = e,~\mu,~n,~p)$ are the equilibrium particle fractions.
The various chemical potential for each particle $i$ is
\begin{equation}\label{se-chempot}
  \mu_i = \frac{\partial \varepsilon/\rho}{\partial Y_i}.
\end{equation}
Through solving Eqs.~\eqref{chemequilb}, \eqref{chargneut}, \eqref{nm-ene}, and \eqref{se-chempot} simultaneously, one can obtain the particle fractions $Y_i$ at a given density $\rho$.
Then the energy density $\varepsilon$ can be easily obtained from Eq.~\eqref{nm-ene}.
The pressure $P$ for the $\beta$-stable nuclear matter is defined as
\begin{equation}\label{nm-pre}
  P=-\frac{\partial (\varepsilon/\rho)}{\partial (1/\rho)}=\rho\frac{\partial \varepsilon}{\partial \rho}-\varepsilon.
\end{equation}

Once the EOS of $\beta$-stable nuclear matter in the form $P(\varepsilon)$ is obtained, the mass and radius of a cold, spherically symmetric, static, and relativistic star can be described by the Tolman-Oppenheimer-Volkov (TOV) equations~\citep{Oppenheimer1939,Tolman1939},
\begin{subequations}\label{ns-toveq}
  \begin{align}
    \frac{dP(r)}{dr}=&\ -\frac{[P(r)+\varepsilon(r)][M(r)+4\pi r^3P(r)]}{r[r-2M(r)]}, \\
	\frac{dM(r)}{dr} =&\ 4\pi r^2\varepsilon(r),
  \end{align}
\end{subequations}
where $P(r)$ is the pressure at neutron star radius $r$, $M(r)$ is the total neutron star mass inside a sphere of radius $r$.
For a given EOS, the TOV equations have the unique solution that depends on the conditions of $\beta$-stable matter at the center, such as the central density or the central pressure.
In addition, to solve the TOV equations, the EOS must cover full regions of the neutron star from the crust to the core.
In the present work, we mainly concentrate on discussing the core region within the RBHF theory.
For the crust, the EOS from the Baym-Pethick-Sutherland (BPS)~\citep{Baym1971} and the Baym-Bethe-Pethick (BBP)~\citep{Baym19712} model is used.

Besides the masses and radii, another important property of neutron star, the tidal deformability $\Lambda$ is defined as
\begin{equation}
  \Lambda = \frac{2}{3}k_2 C^{-5},
\end{equation}
which represents the mass quadrupole moment response of a neutron star to the strong gravitational field induced by its companion~\citep{Damour1992,Hinderer2008,Flanagan2008,Damour2009,Postnikov2010}. $C=M/R$ is the compactness parameter, $M$ and $R$ are the neutron star mass and radius, and $k_2$~is the second love number
\begin{equation}\label{ns-luv}
  \begin{aligned}
    k_2=&\frac{8C^5}{5}(1-2C)^2[2-y_R+2C(y_R-1)]\times\{6C[2-y_R+C(5y_R-8)] \\
	&+4C^3[13-11y_R+C(3y_R-2) +2C^2(1+y_R)] \\
	&+3(1-2C)^2[2-y_R+2C(y_R-1)]\ln(1-2C)\}^{-1},
  \end{aligned}
\end{equation}
where $y_R=y(R)$ can be calculated by solving the following differential equation:
\begin{equation}\label{yRequ}
  r\frac{d y(r)}{dr} + y^2(r)+y(r)F(r) + r^2Q(r)=0,
\end{equation}
with
\begin{subequations}
\begin{align}
  F(r) & = \left[1-\frac{2M(r)}{r}\right]^{-1}\left\{1-4\pi r^2[\varepsilon(r)-P(r)]\right\} ,\\
  \nonumber
  Q(r) & =  \left\{4\pi  \left[5\varepsilon(r)+9P(r)+\frac{\varepsilon(r)+P(r)}{\frac{\partial P}{\partial\varepsilon}(r)}\right]-\frac{6}{r^2}\right\}\times
			\left[1-\frac{2M(r)}{r}\right]^{-1}  \\
  &~~-\left[\frac{2M(r)}{r^2} +2\times4\pi r P(r) \right]^2 \times \left[1-\frac{2M(r)}{r}\right]^{-2} .
\end{align}
\end{subequations}
The differential equation \eqref{yRequ} can be integrated together with the TOV equations with the boundary condition $y(0)=2$.

\section{Results and discussions}\label{Resultsdiscussions}

\subsection{Nuclear matter properties}

\begin{figure}[htbp]
  \centering
  \includegraphics[width=0.8\textwidth]{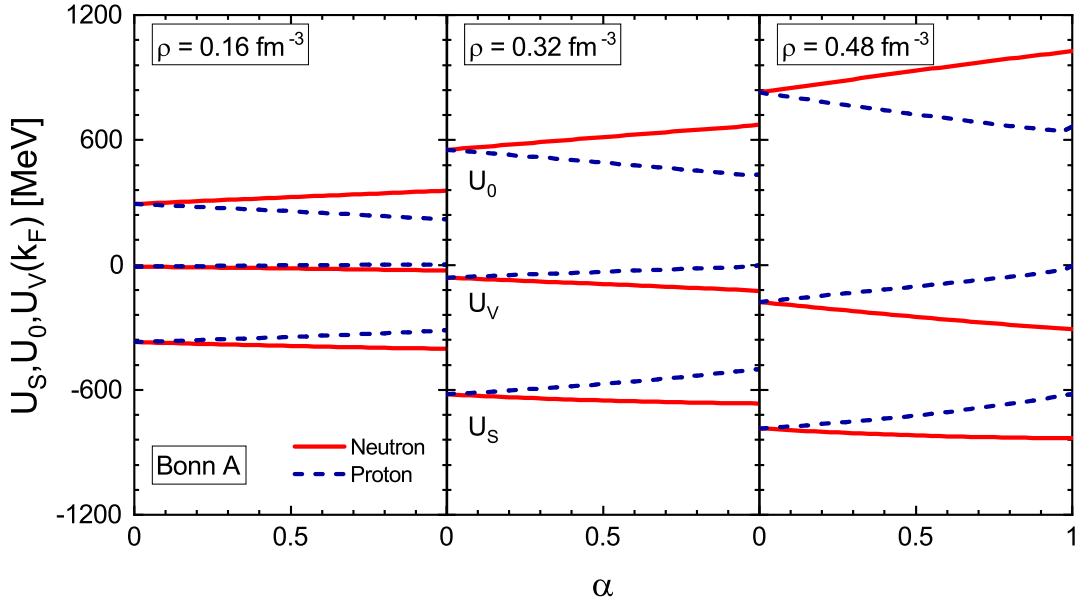}
  \caption{
  Single-particle potentials $U_S$, $U_0$, and $U_V$ at the Fermi momentum $k_F$ as functions of the asymmetric parameter $\alpha$ calculated by the RBHF theory in the full Dirac space.
  The solid (red) and dashed (blue) lines correspond to neutron and proton.
  The total baryon densities are given at $\rho=0.16$~fm$^{-3}$, 0.32~fm$^{-3}$, and 0.48~fm$^{-3}$.
  }
  \label{fig1}
\end{figure}

In ANM, one of the most important questions is the isospin and density dependence of the single-particle potentials~\citep{VanDalen2005_PRL95-022302}, which is related to a wildly-discussed topic concerning the proton-neutron mass splitting.
In Fig.~\ref{fig1}, the isospin and density dependence of single-particle potentials $U_{S,\tau}, U_{0,\tau}$ and $U_{V,\tau}$ at the Fermi momentum $k^\tau_F$ for densities $\rho=0.16~\text{fm}^{-3}$, $0.32~\text{fm}^{-3}$ and $0.48~\text{fm}^{-3}$ are obtained from RBHF calculation in the full Dirac space.
At the density $\rho=0.16~\text{fm}^{-3}$, the values of $U_{V,\tau}$ for both neutron and proton are close to zero.
For the other two components, the RBHF calculation in the full Dirac space gives $U_{S,n}<U_{S,p}$ and $U_{0,n}>U_{0,p}$.
It is also found that with increasing neutron excess, $U_{S,n}$ and $U_{V,n}$ decrease, while $U_{0,n}$ increases.
The case for proton shows an opposite behavior.
In the RMF theory~\citep{Ring1996_PPNP37-193}, the isospin splittings of the scalar and the timelike part of the vector potential can be calculated as~\citep{Ulrych1997_PRC56-1788}
\begin{subequations}\label{equRMF}
	\begin{align}
  		U_{S,n} - U_{S,p} =&\ -2 \frac{g^2_\delta}{m^2_\delta} (\rho_{s,n} - \rho_{s,p}), \\
  		U_{0,n} - U_{0,p} =&\ +2 \frac{g^2_\rho}{m^2_\rho} (\rho_{n} - \rho_{p}).
	\end{align}
\end{subequations}
The parameters $g_i$ and $m_i\ (i=\delta,\rho)$ refer to the meson-nucleon coupling constants and masses of the mesons.
$\rho_{s,\tau}$ is the scalar density.
For ANM, Eqs.~\eqref{equRMF} lead to $U_{S,n}<U_{S,p}$ and $U_{0,n}>U_{0,p}$, which is consistent with the results by the RBHF theory in the full Dirac space.
This consistence underlines the crucial importance for the $\delta$-meson exchange in the RMF theory~\citep{Kubis1997_PLB399-191,Hofmann2001_PRC64-034314,LIU-B2002_PRC65-045201,RocaMaza2011_PRC84-054309}.
Further, since neutron star radii and tidal deformabilities are sensitive to the nuclear matter properties at supra-saturation density~\citep{Fattoyev2018PRL,Zhang2019EPJA,Tong2020PRC}.
We also study the isospin dependence of single-particle potentials at supra-saturation densities $\rho=0.32~\text{fm}^{-3}$ and $0.48~\text{fm}^{-3}$.
The differences between neutron and proton becomes larger when increasing nuclear matter density.
In comparison to the density $\rho=0.16~\text{fm}^{-3}$, the amplitudes of $U_{V,\tau}$ for both neutron and proton are much larger than zero.
However, $U_{V,\tau}$ is neglected in the momentum-independence approximation even at supra-saturation densities.
It means that determining the correct behavior of the isospin dependence of the single-particle potentials at supra-saturation densities is important to describe nuclear matter and neutron star properties.

\begin{figure}[!htbp]
  \centering
  \includegraphics[width=0.5\textwidth]{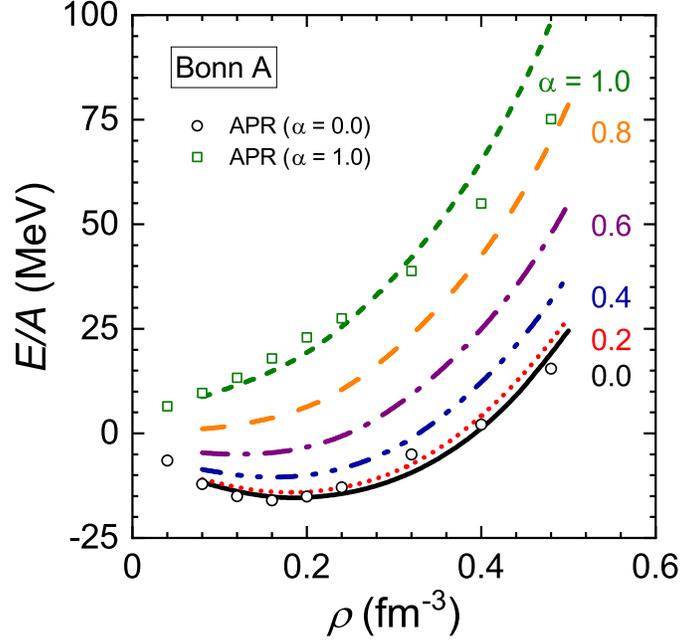}
  \caption{
  Binding energies per nucleon $E/A$ as functions of the density $\rho$ calculated by the RBHF theory in the full Dirac space with the asymmetry parameter $\alpha$ ranging from $0$ to $1$.
  For comparison, the EOS for SNM and pure neutron matter (PNM) obtained with the variational method (APR)~\citep{Akmal1998_PRC58-1804} are also shown with black circles and green squares, respectively.
  }
  \label{fig2}
\end{figure}

From single-particle potentials one can calculate the binding energies per nucleon $E/A$ for ANM.
In Fig.~\ref{fig2}, $E/A$ are depicted as functions of the density $\rho$ with the asymmetry parameter $\alpha$ ranging from $0$ to $1$.
The binding energy and the saturation density become progressively smaller when $\alpha$ increases.
The EOS of ANM exhibits a minimum which disappears for $\alpha\geqslant 0.8$.
For comparison, Fig.~\ref{fig2} also displays the EOS for SNM and PNM obtained with the variational calculations~\citep{Akmal1998_PRC58-1804} using the Argonne $v_{18}$ interaction~\citep{Wiringa1995_PRC51-38} with relativistic boost corrections to the two-nucleon interaction, as well as three-nucleon interactions modeled with the Urbana force~\citep{Pudliner1995_PRL74-4396}.
It can be seen that the results obtained with two \emph{ab initio} methods are comparable.

\begin{table}[htbp]
  \centering
  \caption{Nuclear matter properties (as described in the text) at saturation density $\rho_0$ calculated by the RBHF theory in the full Dirac space, in comparison with the results obtained by the RBHF calculation with the projection method and the momentum-independence approximation (Mom.-ind.app). Results for Bonn potentials A, B, and C are shown. The empirical values are also listed in the last row.}
  \begin{tabular}{ccccccccc}
    \hline\hline

    \multirow{2}{*}{Model} & \multirow{2}{*}{Potential} & $\rho_0$ & $E/A$ & $K_{\infty}$ & $E_{\mathrm{sym}}$ & $L$ & $K_{\mathrm{sym}}$ & $K_{\mathrm{asy}}$   \\
    \multicolumn{2}{c}{} & (fm$^{-3}$) & (MeV) & (MeV) & (MeV) & (MeV) & (MeV) & (MeV)   \\

    \hline

    & A & 0.188 & -15.40 & 258 & 33.1 & 65.2 & -67.3 & -463  \\
    Full Dirac Space & B & 0.164 & -13.36 & 206 & 28.9 & 48.6 & -64.1 & -356 \\
    & C & 0.144 & -12.09 & 150 & 25.8 & 38.0 & -58.1 & -286   \\

    \hline

    & A & 0.179 & -16.15 & 254 & 34.7 & 68.8 & -70.1 & -483  \\
    Projection method & B & 0.161 & -14.60 & 206 & 31.3 & 54.3 & -68.7 & -395  \\
    & C & 0.149 & -13.67 & 157 & 29.1 & 45.5 & -65.3 & -339   \\

    \hline

    & A & 0.178 & -15.36 & 273 & 33.2 & 67.3 & -65.7 & -469  \\
    Mom.-ind.app & B & 0.162 & -13.45 & 233 & 29.7 & 55.0 & -63.2 & -393  \\
    & C & 0.150 & -12.14 & 201 & 27.1 & 46.9 & -59.6 & -341  \\

    \hline

    \multirow{2}{*}{Empirical} & & 0.166 & -16 & 240 & 32 & 88 &  & \\
    & & $\pm$0.018 & $\pm$1 & $\pm$20 & $\pm$2 & $\pm$25 &  & \\

    \hline\hline
  \end{tabular}
  \label{tab1}
\end{table}

In Table~\ref{tab1}, we summarize our results for the properties of nuclear matter from RBHF theory.
In the second row, we show our results calculated by the RBHF theory in the full Dirac space.
The results obtained by the projection method~\citep{GROSSBOELTING1999} with the $ps$ representation for the subtracted $T$ matrix~\citep{VanDalen2004_NPA744-227} are shown in the third row.
The results obtained by the momentum-independence approximation are also shown in the fourth row, the scalar potential and the time component of the vector potential are extracted directly from the single-particle potential energy at two casually selected momenta, $0.5k_F^\tau$ and $k_F^\tau$.
The one-boson-exchange potential (OBEP) is used for the realistic $NN$ interactions, which is defined as a sum of one-particle-exchange amplitudes of six non-strange bosons with given spin-parity properties~\citep{Machleidt1989}.
The coupling strengths and form-factor parameters are determined by fitting to the $NN$ scattering data and deuteron properties.
Three different parametrizations of OBEP were constructed for the application to relativistic nuclear structure physics, which were denoted by Bonn A, B, and C.
The three potentials differ essentially in the $\pi NN$ form-factor parameter.
Since the nuclear tensor force is essentially provided by the pion, the main difference among the three potentials is the strength of the tensor force which can be reflected in their predictions for $D$-state probability of the deuteron ($P_D = 4.5\%,~5.1\%,~5.5\%$ for Bonn A, B, C, respectively)~\citep{Machleidt1987}.
As is well known, the strength of the tensor force determines the location of the nuclear matter saturation point~\citep{Coester1970,Brockmann1990}.
Compared to potential Bonn B and C, all three methods by using potential Bonn A with the weakest tensor force can reproduce the empirical value~\citep{Bethe1971,Sprung1972} for saturation density $\rho_0$ and binding energy $E/A$.
With the potential Bonn A, the incompressibility of SNM at saturation density is 258~MeV for the RBHF calculation in the full Dirac space and 254~MeV in the projection method, which agree well with the commonly accepted value of 240~$\pm$~20~MeV~\citep{Garg2007,Garg2018}.
As for ANM, the symmetry energy $E_{\mathrm{sym}}$ and the slope parameter $L$ in the full Dirac space with potential Bonn A are $33.1$~MeV and $65.2$~MeV, which are smaller than the values suggested in the projection method and the momentum-independence approximation.
This implies that the full Dirac space predicts the softest symmetry energy at higher densities.

\begin{figure}[!htbp]
  \centering
  \includegraphics[width=0.9\textwidth]{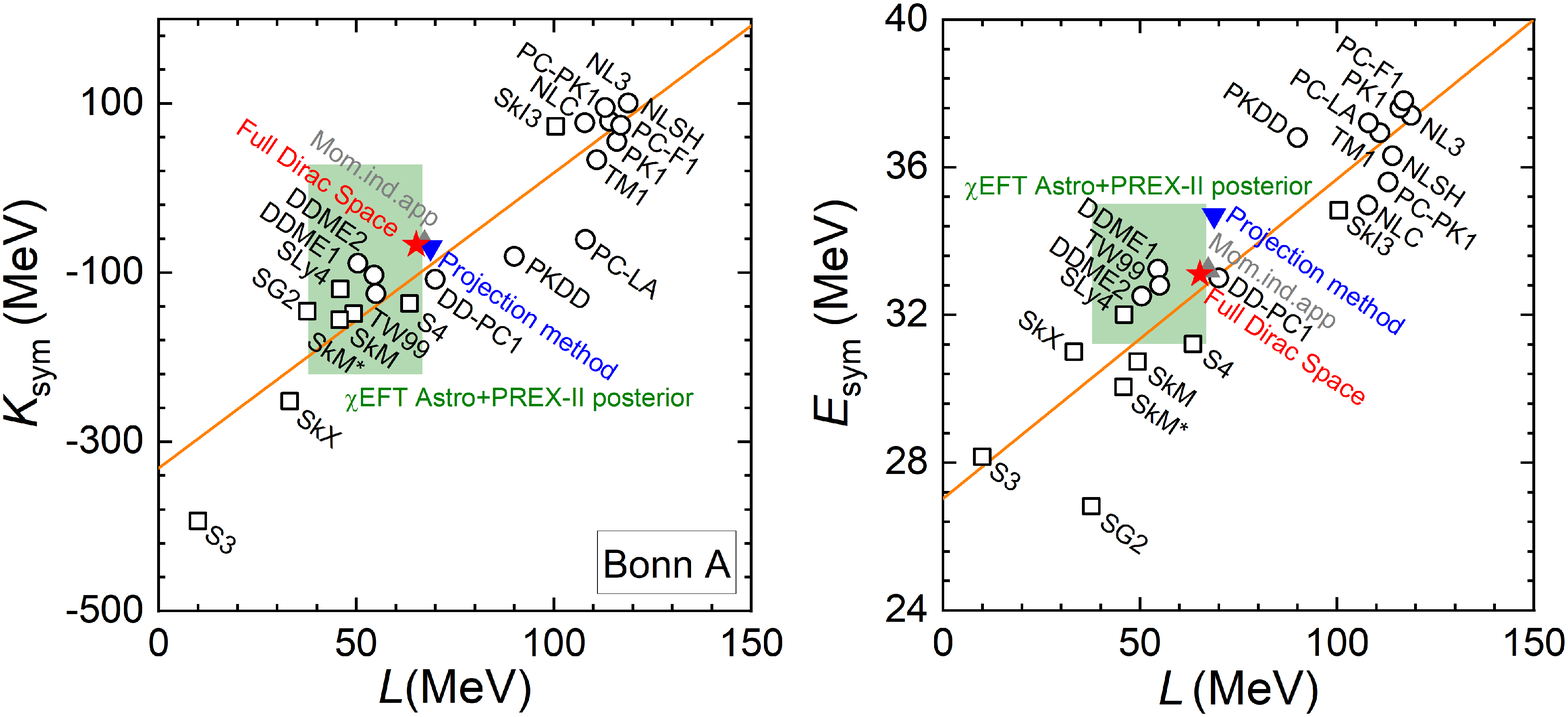}
  \caption{
  $K_{\mathrm{sym}}$ (left panel) and
  $E_{\mathrm{sym}}$ (right panel) as functions of $L$ calculated by RBHF theory using the potential Bonn A in the full Dirac space (red star), in comparison with results obtained by the projection method (blue down-pointing triangle), the momentum-independence approximation (gray up-pointing triangle) and various density functionals (circles and squares)~\citep{Sagawa2007,ZhaoLiYaoEtAl2010}.
  The shaded regions denote a recent constraint on $E_{\mathrm{sym}}$, $L$ and $K_{\mathrm{sym}}$ by combining astrophysical data with PREX-II and chiral effective field theory~\citep{Essick2021PRL}.
  The solid orange line is a linear fit to the results of density functionals.
  }
  \label{fig3}
\end{figure}

In several previous works, slope parameter $L$ and curvature parameter $K_{\mathrm{sym}}$ of symmetry energy $E_{\mathrm{sym}}$, as well as their correlations, were shown to have significant effects on the neutron star properties, e.g., the composition of the core~\citep{Lattimer1991PRL}, crust-core transition density and pressure~\citep[e.g.,][]{Vidana2009PRC,NaiBoZhang2018APJ,BaoAnLi2020PRC}.
At the saturation density, $E_{\mathrm{sym}}$ has been well constrained by recent terrestrial nuclear experiments and astrophysical observations.
However, $L$ and $K_{\mathrm{sym}}$ still remain relatively large uncertainties which characterize the $E_{\mathrm{sym}}$ at supra-saturation densities.
In Fig.~\ref{fig3}, $K_{\mathrm{sym}}$, $L$ and $E_{\mathrm{sym}}$ are calculated by a series of DFTs and the \emph{ab initio} RBHF theory.
The DFTs are adopted from nonrelativistic models (e.g., SLy4 and those starting with S)~\citep{Dutra2012}, to relativistic models (e.g., PK1, NL3, DD-ME)~\citep{Long2006,Dutra2014}.
\emph{Ab initio} calculations have been carried out with the RBHF theory using the potential Bonn A in the full Dirac space, together with results obtained by the projection method and the momentum-independence approximation.
$E_{\mathrm{sym}}$ and $K_{\mathrm{sym}}$ obtained from DFTs exhibit a positive correlation with $L$, and the results of RBHF theory are all in good agreement with this correlation.
We also perform a linear fitting to this correlation,
\begin{subequations}\label{linearfit}
	\begin{align}
  		K_{\mathrm{sym}} =&\ -331.32+3.49L, \\
  		E_{\mathrm{sym}} =&\ 27.03+0.09L,
	\end{align}
\end{subequations}
with the correlation coefficients $r = 0.93$ and $r = 0.92$.
The results of fitting are depicted with the solid orange line.
Our results are close to $K_{\mathrm{sym}} = -307.862 + 3.292L$ with $r=0.972$ given in Ref.~\citep{DongPRC2012} which used several different DFTs from the present work.
It should be mentioned that $K_{\mathrm{sym}}$ predicted by RBHF theory are negative values, i.e., -67.3~MeV in full Dirac space, -70.1~MeV in the projection method and -65.7~MeV in the momentum-independence approximation.
In addition, the shaded regions in Fig.~\ref{fig3} denote a recent constraint on $E_{\mathrm{sym}}$, $L$,  and $K_{\mathrm{sym}}$ by combining astrophysical data with PREX-II and chiral effective field theory~\citep{Essick2021PRL}.
It can be seen that the results in the full Dirac space located inside the region labeled by $\chi$EFT Astro+PREX-II posterior.
Our results are also consistent with other recent works, such as $K_{\mathrm{sym}}=-120_{-100}^{+80}$~MeV at $68\%$ confidence level from a Bayesian analysis of neutron star properties~\citep{WenJieXie2020APJ} and
$K_{\mathrm{sym}}=-209_{-182}^{+270}$~MeV at $95\%$ confidence level from combining the neutron skin data and the pure neutron matter EOS from chiral effective field theory calculations~\citep{Newton2021PRC}.

\subsection{Neutron star properties}

\begin{figure}[!htbp]
  \centering
  \includegraphics[width=0.9\textwidth]{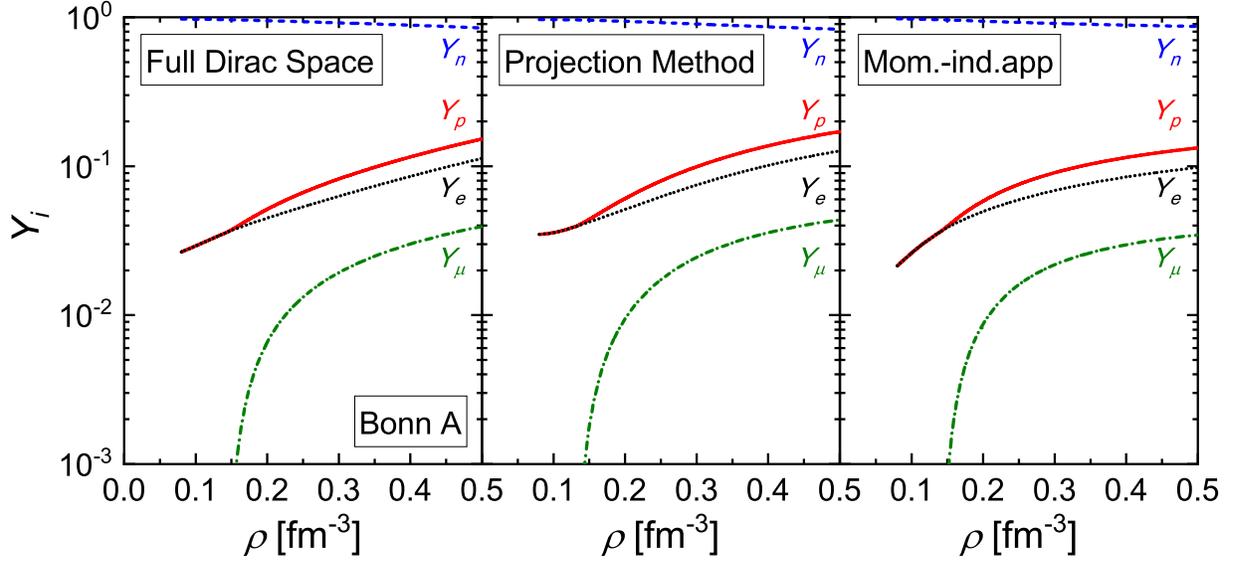}
  \caption{Particle fractions $Y_i~(i=n,p,e,\mu)$ of $\beta$-stable matter as a function of the baryon number density. The results are calculated by RBHF theory using the potential Bonn A in the full Dirac space (left panel), together with the projection method (middle panel), and the momentum-independence approximation (right panel).
  }
  \label{fig4}
\end{figure}

On the basis of the Eqs.~\eqref{chemequilb}, \eqref{chargneut}, \eqref{nm-ene}, and \eqref{se-chempot}, the particle fractions $Y_i~(i=n,p,e,\mu)$ for $\beta$-stable matter in Fig.~\ref{fig4} are calculated by RBHF theory using the potential Bonn A in the full Dirac space, together with the projection method, and the momentum-independence approximation.
The neutron fractions $Y_n$ decrease gradually with increasing $\rho$ in three methods.
In contrast, the values of $Y_p$, $Y_e$, and $Y_\mu$ grow steadily.
According to the energetic constraints~\citep{Wiringa1988PRC}, muons only emerge at higher densities, and the onset densities for muons from three methods are 0.144~fm$^{-3}$, 0.132~fm$^{-3}$ and 0.142~fm$^{-3}$.
The muons appear latest in the full Dirac space, since it has the softest symmetry energy and generates the smallest proton fraction.
We also observe that the behavior of particle fractions in the full Dirac space is qualitatively similar to the one seen in other two methods.

The direct URCA (DURCA) process
\begin{equation}
  n \rightarrow p + e^- + \bar{\nu}_e, \quad p \rightarrow n + e^+ + \nu_e,
\end{equation}
is one of the scenario of neutron star cooling~\citep{Pethick1992RMP,Prakash1992APJ,Prakash1994PR}.
This process is allowed when the momenta for nucleons and leptons can be conserved, which can predict the threshold density of the DURCA process.
It has been pointed out in Ref.~\citep{Lattimer2004} that the $\beta$-stable matter with a proton-to-neutron ratio in excess of $1/8$, or the proton fraction $Y_p \geqslant 1/9$ can be cooled by the DURCA process.
As for the results from RBHF theory, the DURCA process happens after muon occurs for three methods.
The threshold densities are calculated as $\rho=0.432$~fm$^{-3}$ and $Y_p=0.127$ in the full Dirac space, $\rho=0.372$~fm$^{-3}$ and $Y_p=0.126$ in the projection method, and $\rho=0.464$~fm$^{-3}$ and $Y_p=0.127$ in the momentum-independence approximation.
These results from three methods are consistent with the astronomical observations, which does not allow the DURCA process to be too early.

\begin{figure}[!htbp]
  \centering
  \includegraphics[width=0.5\textwidth]{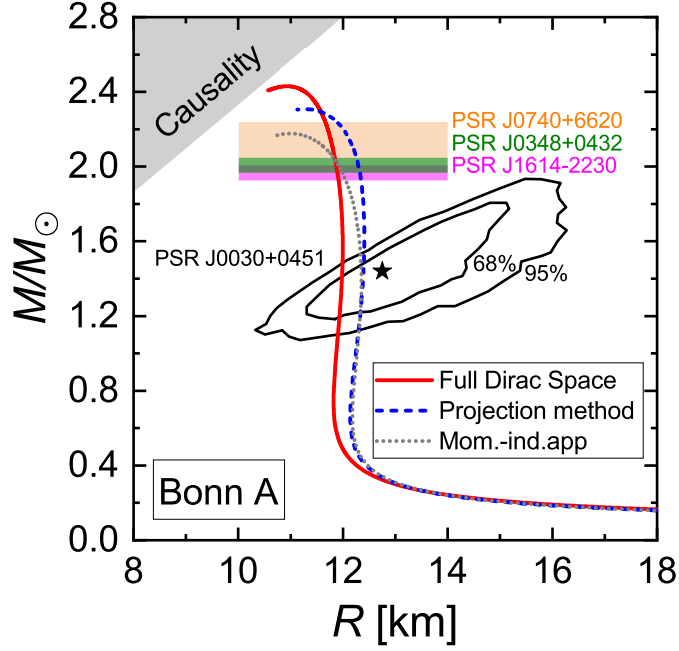}
  \caption{
  Neutron star mass-radius relations obtained by RBHF theory using the potential Bonn A in the full Dirac space (red solid line), together with the projection method (blue dashed line), and the momentum-independence approximation (gray dotted line).
  The shaded areas represent the constraints from astronomical observables for the massive neutron star~\citep{demorest2010,Fonseca2016APJ,antoniadis2013,cromartie2020}.
  The inner and outer contours indicate the allowed area of mass and radius of neutron stars by NICER's analysis of PSR~J0030+451~\citep{Miller2019}.
  The excluded causality region is also shown~\citep{LATTIMER2007PR}.
  }
  \label{fig5}
\end{figure}

Based on the EOS from RBHF theory, the mass-radius relations of neutron stars can be calculated from the TOV equation.
In Fig.~\ref{fig5}, the neutron star mass-radius relations obtained are shown from the RBHF theory in the full Dirac space with potential Bonn A, together with the projection method, and the momentum-independence approximation.
The radii $R_{1.4M_\odot}$ of a $1.4M_{\odot}$ neutron star from three methods are 11.97~km, 12.38~km, and 12.35~km, respectively.
The prediction from the full Dirac space leads to the smallest radius for $R_{1.4M_\odot}$.
The relatively small neutron star radius suggested by full Dirac space implies that the symmetry energy at higher densities is soft.
It should be noted that soft EOSs, which produce small values of radii for fixed $M$ are favored by the GW170817~\citep{Fattoyev2018PRL}.
It can be found that our results for neutron star radii are also consistent with other works, such as $R_{1.4M_{\odot}}\leqslant 13.76$~km~\citep{Fattoyev2018PRL}, $R_{1.4M_{\odot}}\leqslant 13.6$~km~\citep{Annala2018}, 12.00~km $\leqslant R_{1.4M_{\odot}}\leqslant 13.45$~km~\citep{Most2018} and 9.0~km $\leqslant R_{1.4M_{\odot}}\leqslant 13.6$~km~\citep{Tews2018} from the tidal deformability~\citep{Abbott2017PRL,Abbott2019PRX}, $R_{1.4M_{\odot}}=10.9_{-1.5}^{+1.9}$~km from the Bayesian approach~\citep{kumar2019} and the recent constraints by NICER~\citep{Miller2019} for the mass and radius of PSR~J0030+0451, i.e., mass $1.44_{-0.14}^{+0.15}M_{\odot}$ with radius $13.02_{-1.06}^{+1.24}$~km.
The $68\%$ and $95\%$ contours of the joint probability density distribution of mass and radius from the NICER analysis are also shown.

The neutron star matter is only considered to be composed of nucleons and leptons here, the central densities of massive neutron stars in this work can reach $5$ to $7$ times the nuclear saturation density $\rho_0=0.16$~fm$^{-3}$ of neutrons and protons found in laboratory nuclei, which is far away from the region $\rho\leqslant 0.57$~fm$^{-3}$ where RBHF theory in the full Dirac space is applicable~\citep{SiboWang2022arxiv}.
Therefore, for RBHF theory in the full Dirac space, we also follow the strategy in Refs.~\citep{Rhoades1974PRL,Gandolfi2012PRC,SiboWang2022arxiv}. The maximally stiff or causal EOS given by $p(\varepsilon)= c^2\varepsilon - \varepsilon_c$ is used to replace the EOS for neutron star matter above a critical density $\rho_c$, where $p$ is the pressure, $\varepsilon$ is the energy density, $c$ is the speed of light, and $\varepsilon_c$ is a constant.
This EOS predicts the most rapid increase of $p$ with $\varepsilon$ without violating causality limit and the upper bound on the maximum mass of the neutron star.
In our calculations, $\rho_c$ is given as 0.57~fm$^{-3}$ and $\varepsilon_c$ is determined by ensuring that the energy density is continuous between the low- and high-density EOS.
The upper bound maximum masses of neutron stars are predicted as 2.43$M_\odot$ in the full Dirac space, which is still smaller than the compact object with a mass of 2.50-2.67$M_\odot$ by LIGO Scientific and Virgo collaborations in GW190814~\citep{Abbott2020APJ}.
Moreover, the maximum masses are predicted as 2.31$M_\odot$ and 2.18$M_\odot$ from the projection method and the momentum-independence approximation, which are consistent with the measurements of massive neutron stars, such as PSR~J1614-2230~\citep{demorest2010,Fonseca2016APJ}, PSR~J0348+0432~\citep{antoniadis2013}, and PSR~J0740+6620~\citep{cromartie2020}.

\begin{figure}[!htbp]
  \centering
  \includegraphics[width=0.6\textwidth]{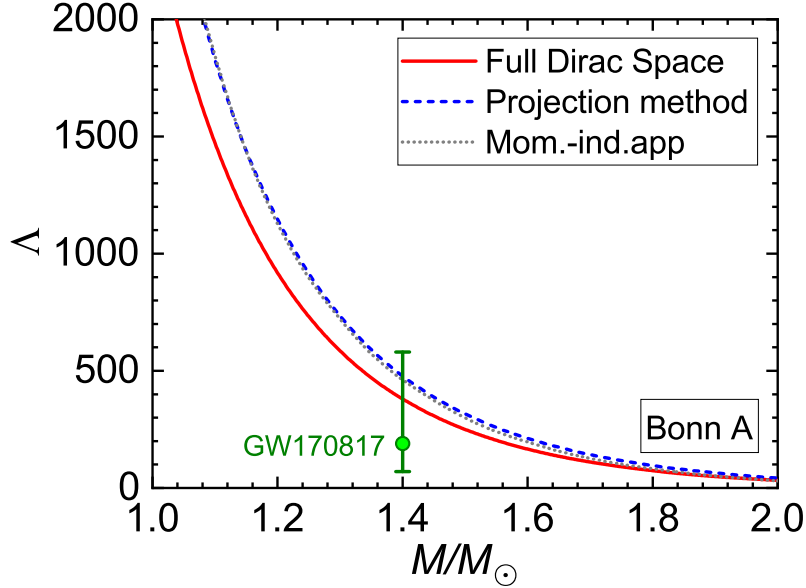}
  \caption{Tidal deformabilities $\Lambda$ as functions of neutron star mass $M$ obtained by RBHF theory using the potential Bonn A in the full Dirac space (red solid line), together with the projection method (blue dashed line), and the momentum-independence approximation (gray dotted line).
  Constraints from the BNS merger event GW170817 for the tidal deformability, $\Lambda_{1.4M_\odot}=190_{-120}^{+390}$~\citep{Abbott2018PRL}, is also shown.
  }
  \label{fig6}
\end{figure}

In the multimessenger era, another important constraint of the neutron star is the tidal deformability $\Lambda$.
In Fig.~\ref{fig6}, the tidal deformabilities $\Lambda$ as functions of neutron star mass $M$ from RBHF theory with potential Bonn A in the full Dirac space, the projection method, and the momentum-independence approximation are shown.
These tidal deformabilities decrease with increasing the neutron star mass.
For a given neutron star mass, the method which has a softer symmetry energy yield smaller neutron star radii and correspondingly smaller tidal deformabilities.
The tidal deformabilities of a $1.4M_{\odot}$ neutron star from three methods are predicted as $\Lambda_{1.4M_\odot}$ = 376, 473, and 459, respectively.
It can be found that the tidal deformabilities from the full Dirac space are significantly lower than those from the projection method and the momentum-independence approximation.
The initial estimation for tidal deformability $\Lambda_{1.4M_\odot}$ has an upper bound $\Lambda_{1.4M_\odot}<800$~\citep{Abbott2017PRL} from the observation of BNS merger event GW170817.
Then a revised analysis from LIGO and Virgo collaborations gives $\Lambda_{1.4M_\odot}=190_{-120}^{+390}$~\citep{Abbott2018PRL}. It is important to underscore that the results from three
methods are located in these regions, especially that from the full Dirac space, which is the closest one to the central values of the GW170817 shown as the circle symbol.

\begin{table}[htbp]
  \centering
  \caption{Neutron star properties calculated by the RBHF theory in the full Dirac space, together with the results obtained by the RBHF calculation with the projection method and the momentum-independence approximation.
  Results for Bonn potentials A, B, and C are shown.}
  \begin{tabular}{cccccccccc}
    \hline\hline

    \multirow{2}{*}{Method} & \multirow{2}{*}{Potential}  & $\rho_{\rm{DURCA}}$ & \multirow{2}{*}{$Y_{p,\rm{DURCA}}$} & $M_{\rm{DURCA}}$
    & $R_{1.4M_\odot}$ & $\rho_{1.4M_\odot}$ &  \multirow{2}{*}{$\Lambda_{1.4M_\odot}$} & $M_{\rm{max}}$ & $R_{\rm{max}}$   \\
    & &(fm$^{-3}$)& &($M_\odot$)&(km)&(fm$^{-3}$)& &($M_\odot$)&(km) \\

    \hline

    & A & 0.43 & 0.13 & 1.28 & 11.97 & 0.46 & 376 & 2.43 & 10.93   \\
    Full Dirac Space & B & 0.48 & 0.13 & 1.55 & 12.13 & 0.45 & 405 & 2.43 & 10.98   \\
    & C & 0.52 & 0.13 & 1.78 & 12.27 & 0.43 & 433 & 2.44 & 11.04   \\

    \hline

    & A & 0.37 & 0.13 & 1.16 & 12.38 & 0.42 & 473 & 2.31 & 11.27  \\
    Projection method & B & 0.41 & 0.13 & 1.41 & 12.53 & 0.41 & 505 & 2.31 & 11.34   \\
    & C & 0.45 & 0.13 & 1.64 & 12.64 & 0.40 & 532 & 2.32 & 11.42  \\

    \hline

    & A & 0.46 & 0.13 & 1.52 & 12.35 & 0.43 & 459 & 2.18 & 10.99   \\
    Mom.-ind.app & B & 0.54 & 0.13 & 1.79 & 12.49 & 0.43 & 484 & 2.18 & 11.05   \\
    & C & 0.63 & 0.13 & 1.97 & 12.59 & 0.42 & 508 & 2.18 & 11.10   \\

    \hline\hline
  \end{tabular}
  \label{tab2}
\end{table}

In Table~\ref{tab2}, our results for neutron star properties, such as threshold of DURCA process, the radii and tidal deformability for $1.4M_\odot$, and central densities, are obtained by the RBHF theory in the full Dirac space, the projection method and the momentum-independence approximation with the potential Bonn A, B, C.
It was pointed out that the strength of tensor force plays an important role in the investigation of neutron star properties by the RBHF theory in the momentum-independence approximation~\citep{Wang2020}.
Therefore, it is interesting to investigate the effect of tensor force on the neutron star properties in the RBHF theory, especially in the full Dirac space.
The threshold densities for DURCA process are listed in the third column.
As for the full Dirac space, $\rho_{\rm{DURCA}}$ from potential Bonn A, B, C are $0.43$~fm$^{-3}$, $0.48$~fm$^{-3}$ and 0.52~fm$^{-3}$, with the same DURCA proton fractions $Y_p=0.13$.
An inverse correlation can be found between the DURCA threshold density and the symmetry energy, since Bonn A
generates the largest symmetry energy at a given baryon
density with the weakest tensor force.
The critical mass for DURCA process are also listed in the fifth column, $M_{\rm{DURCA}}$ from potential Bonn A, B, C are $1.28 M_\odot$, $1.55 M_\odot$, and $1.78 M_\odot$ in the full Dirac space. Recently, the analysis of a neutron star in the transient system MXB~1659-29 has founded that the critical mass is around $1.6M_\odot$~\citep{BrownPRL2018}, and we found that the results from Bonn B is the closest one to the current astrophysical observations.
The neutron star radii and tidal deformabilties at 1.4$M_\odot$ from the full Dirac space are shown in the sixth and the eighth column, which are given as $R_{1.4M_\odot}$ = 11.97~km and $\Lambda_{1.4M_\odot}=376$ for Bonn A, $R_{1.4M_\odot}$ = 12.13~km and $\Lambda_{1.4M_\odot}=405$ for Bonn B, and $R_{1.4M_\odot}$ = 12.27~km and $\Lambda_{1.4M_\odot}=433$ for Bonn C.
We found that $R_{1.4M_\odot}$ and $\Lambda_{1.4M_\odot}$ predicted by Bonn A are smaller than those from Bonn B and C.
This indicates that Bonn A with the lowest strength of tensor force has the softest EOS and Bonn C has the stiffest EOS.
In the ninth column, the upper bound on the maximum masses of neutron stars from the full Dirac space are shown as 2.43$M_\odot$, 2.43$M_\odot$, 2.44$M_\odot$ with Bonn A, B, C.
The difference among the three kinds of results is smaller than 0.01$M_\odot$.
This is reasonable since neutrons dominate the nucleonic component of the maximum masses of neutron stars while the main difference between the three Bonn potentials is in their tensor force strength in the $(T=0)$~$^3S_1$-$^3D_1$ state that does not contribute to the $(T=1)$ neutron-neutron state.

\section{Summary}\label{summary}

The single-particle potential can be used to describe the quasi-particle properties of a nucleon inside a strongly interacting nuclear matter system.
However, because the incompleteness of the Dirac space, different approximations have been introduced to extract the momentum and isospin dependence of the single-particle potential in the RBHF theory.
In contrast to the RBHF calculations in the Dirac space with PESs only, the single-particle potential can be determined in a unique way from RBHF theory in the full Dirac space.
The scalar component for neutron is found to be smaller than that for proton at and above nuclear saturation density, i.e., $U_{S,n}<U_{S,p}$, which leads to the neutron-proton Dirac mass splitting $M^*_{D,n}<M^*_{D,p}$.
As for the timelike component of the vector potential, the RBHF theory in the full Dirac space gives $U_{0,n}>U_{0,p}$.
At higher densities, the amplitudes of $U_{V,\tau}$ for both neutron and proton are much larger than zero, it means $U_{V,\tau}$ at higher densities can not be neglected.
The saturation properties of symmetric and asymmetric nuclear matter are systematically investigated based on the Bonn potentials.
The symmetry energy $E_{\mathrm{sym}}$ and its slope parameter $L$ at the saturation density are $33.1$ and $65.2$ MeV for potential Bonn A, which are consistent with the empirical and experimental values.
Especially, the slope parameter $L$ and the curvature parameter $K_{\mathrm{sym}}$ predicted in the full Dirac space agree well with the recent results from $\chi$EFT Astro+PREX-II posterior.

To further study the importance of the calculations in the full Dirac space at supra-saturation densities, the neutron star properties are studied with the latest RBHF theory in the full Dirac space.
The threshold densities of the DURCA process are calculated as $\rho=0.43$~fm$^{-3}$ and $Y_p=0.13$ in the full Dirac space.
These results are consistent with the astronomical observations, which does not allow the DURCA process to be too early.
The radii of a $1.4M_\odot$ neutron star are calculated as $R_{1.4M_\odot}=11.97,~12.13,~12.27$~km, and their tidal deformabilities are $\Lambda_{1.4M_\odot}=376,~405,~433$ for potential Bonn A,~B,~C.
Comparing the results obtained with the projection method and the momentum-independence approximation, the RBHF calculations in the full Dirac space predict the softest EOS which would be more favored by GW170817.
Especially, the tidal deformability from the full Dirac space is the closest one to the central values of the GW170817 from a revised analysis by LIGO and Virgo collaborations.
Among the potential Bonn A,~B,~C, it is also found that RBHF calculations with the potential Bonn A has the softest EOS and potential Bonn C has the stiffest EOS since the former one has the lowest strength of tensor force.
Furthermore, the results from RBHF theory are consistent with the recent astronomical observations of massive neutron stars and simultaneous mass-radius measurement.


\begin{acknowledgments}
S.W. thanks Qiang Zhao for helpful discussions.
This work was partly supported by the National Key R\&D Program of China under Contracts No. 2017YFE0116700, the National Natural Science Foundation of China (NSFC) under Grants 12147102, the Fundamental Research Funds for the Central Universities under Grants No. 2020CDJQY-Z003 and No. 2021CDJZYJH-003, the MOST-RIKEN Joint Project “Ab initio investigation in nuclear physics”.
Part of this work was achieved by using the supercomputer OCTOPUS at the Cybermedia Center, Osaka University under the support of Research Center for Nuclear Physics of Osaka University.
\end{acknowledgments}


\bibliography{reference}{}

\begin{thebibliography}{}
\expandafter\ifx\csname natexlab\endcsname\relax\def\natexlab#1{#1}\fi
\providecommand{\url}[1]{\href{#1}{#1}}
\providecommand{\dodoi}[1]{doi:~\href{http://doi.org/#1}{\nolinkurl{#1}}}
\providecommand{\doeprint}[1]{\href{http://ascl.net/#1}{\nolinkurl{http://ascl.net/#1}}}
\providecommand{\doarXiv}[1]{\href{https://arxiv.org/abs/#1}{\nolinkurl{https://arxiv.org/abs/#1}}}

\bibitem[{Abbott {et~al.}(2017)Abbott, Abbott, Abbott, Acernese, Ackley, Adams,
  Adams, Addesso, Adhikari, Adya, Affeldt, Afrough, Agarwal, Agathos, Agatsuma,
  Aggarwal, Aguiar, Aiello, Ain, Ajith, Allen, Allen, Allocca, Altin, Amato,
  Ananyeva, Anderson, Anderson, Angelova, Antier, Appert, Arai, Araya, Areeda,
  Arnaud, Arun, Ascenzi, Ashton, Ast, Aston, Astone, Atallah, Aufmuth, Aulbert,
  AultONeal, Austin, Avila-Alvarez, Babak, Bacon, Bader, Bae, Bailes, Baker,
  Baldaccini, Ballardin, Ballmer, Banagiri, Barayoga, Barclay, Barish, Barker,
  Barkett, Barone, Barr, Barsotti, Barsuglia, Barta, Barthelmy, Bartlett,
  Bartos, Bassiri, Basti, Batch, Bawaj, Bayley, Bazzan, B\'ecsy, Beer, Bejger,
  Belahcene, Bell, Berger, Bergmann, Bernuzzi, Bero, Berry, Bersanetti,
  Bertolini, Betzwieser, Bhagwat, Bhandare, Bilenko, Billingsley, Billman,
  Birch, Birney, Birnholtz, Biscans, Biscoveanu, Bisht, Bitossi, Biwer,
  Bizouard, Blackburn, Blackman, Blair, Blair, Blair, Bloemen, Bock, Bode,
  Boer, Bogaert, Bohe, Bondu, Bonilla, Bonnand, Boom, Bork, Boschi, Bose,
  Bossie, Bouffanais, Bozzi, Bradaschia, Brady, Branchesi, Brau, Briant,
  Brillet, Brinkmann, Brisson, Brockill, Broida, Brooks, Brown, Brown, Brunett,
  Buchanan, Buikema, Bulik, Bulten, Buonanno, Buskulic, Buy, Byer, Cabero,
  Cadonati, Cagnoli, Cahillane, Calder\'on~Bustillo, Callister, Calloni, Camp,
  Canepa, Canizares, Cannon, Cao, Cao, Capano, Capocasa, Carbognani, Caride,
  Carney, Carullo, Casanueva~Diaz, Casentini, Caudill, Cavagli\`a, Cavalier,
  Cavalieri, Cella, Cepeda, Cerd\'a-Dur\'an, Cerretani, Cesarini, Chamberlin,
  Chan, Chao, Charlton, Chase, Chassande-Mottin, Chatterjee, Chatziioannou,
  Cheeseboro, Chen, Chen, Chen, Cheng, Chia, Chincarini, Chiummo, Chmiel, Cho,
  Cho, Chow, Christensen, Chu, Chua, Chua, Chung, Chung, Ciani, Ciolfi,
  Cirelli, Cirone, Clara, Clark, Clearwater, Cleva, Cocchieri, Coccia, Cohadon,
  Cohen, Colla, Collette, Cominsky, Constancio, Conti, Cooper, Corban, Corbitt,
  Cordero-Carri\'on, Corley, Cornish, Corsi, Cortese, Costa, Coughlin,
  Coughlin, Coulon, Countryman, Couvares, Covas, Cowan, Coward, Cowart, Coyne,
  Coyne, Creighton, Creighton, Cripe, Crowder, Cullen, Cumming, Cunningham,
  Cuoco, Dal~Canton, D\'alya, Danilishin, D'Antonio, Danzmann, Dasgupta,
  Da~Silva~Costa, Dattilo, Dave, Davier, Davis, Daw, Day, De, DeBra, Degallaix,
  De~Laurentis, Del\'eglise, Del~Pozzo, Demos, Denker, Dent, De~Pietri,
  Dergachev, De~Rosa, DeRosa, De~Rossi, DeSalvo, de~Varona, Devenson,
  Dhurandhar, D\'{\i}az, Dietrich, Di~Fiore, Di~Giovanni, Di~Girolamo,
  Di~Lieto, Di~Pace, Di~Palma, Di~Renzo, Doctor, Dolique, Donovan, Dooley,
  Doravari, Dorrington, Douglas, Dovale~\'Alvarez, Downes, Drago,
  Dreissigacker, Driggers, Du, Ducrot, Dudi, Dupej, Dwyer, Edo, Edwards,
  Effler, Eggenstein, Ehrens, Eichholz, Eikenberry, Eisenstein, Essick,
  Estevez, Etienne, Etzel, Evans, Evans, Factourovich, Fafone, Fair, Fairhurst,
  Fan, Farinon, Farr, Farr, Fauchon-Jones, Favata, Fays, Fee, Fehrmann, Feicht,
  Fejer, Fernandez-Galiana, Ferrante, Ferreira, Ferrini, Fidecaro, Finstad,
  Fiori, Fiorucci, Fishbach, Fisher, Fitz-Axen, Flaminio, Fletcher, Fong, Font,
  Forsyth, Forsyth, Fournier, Frasca, Frasconi, Frei, Freise, Frey, Frey,
  Fries, Fritschel, Frolov, Fulda, Fyffe, Gabbard, Gadre, Gaebel, Gair,
  Gammaitoni, Ganija, Gaonkar, Garcia-Quiros, Garufi, Gateley, Gaudio, Gaur,
  Gayathri, Gehrels, Gemme, Genin, Gennai, George, George, Gergely, Germain,
  Ghonge, Ghosh, Ghosh, Ghosh, Giaime, Giardina, Giazotto, Gill, Glover, Goetz,
  Goetz, Gomes, Goncharov, Gonz\'alez, Gonzalez~Castro, Gopakumar, Gorodetsky,
  Gossan, Gosselin, Gouaty, Grado, Graef, Granata, Grant, Gras, Gray, Greco,
  Green, Gretarsson, Groot, Grote, Grunewald, Gruning, Guidi, Guo, Gupta,
  Gupta, Gushwa, Gustafson, Gustafson, Halim, Hall, Hall, Hamilton, Hammond,
  Haney, Hanke, Hanks, Hanna, Hannam, Hannuksela, Hanson, Hardwick, Harms,
  Harry, Harry, Hart, Haster, Haughian, Healy, Heidmann, Heintze, Heitmann,
  Hello, Hemming, Hendry, Heng, Hennig, Heptonstall, Heurs, Hild, Hinderer, Ho,
  Hoak, Hofman, Holt, Holz, Hopkins, Horst, Hough, Houston, Howell, Hreibi, Hu,
  Huerta, Huet, Hughey, Husa, Huttner, Huynh-Dinh, Indik, Inta, Intini, Isa,
  Isac, Isi, Iyer, Izumi, Jacqmin, Jani, Jaranowski, Jawahar,
  Jim\'enez-Forteza, Johnson, Johnson-McDaniel, Jones, Jones, Jonker, Ju,
  Junker, Kalaghatgi, Kalogera, Kamai, Kandhasamy, Kang, Kanner, Kapadia,
  Karki, Karvinen, Kasprzack, Kastaun, Katolik, Katsavounidis, Katzman, Kaufer,
  Kawabe, K\'ef\'elian, Keitel, Kemball, Kennedy, Kent, Key, Khalili, Khan,
  Khan, Khan, Khazanov, Kijbunchoo, Kim, Kim, Kim, Kim, Kim, Kim, Kimbrell,
  King, King, Kinley-Hanlon, Kirchhoff, Kissel, Kleybolte, Klimenko, Knowles,
  Koch, Koehlenbeck, Koley, Kondrashov, Kontos, Korobko, Korth, Kowalska,
  Kozak, Kr\"amer, Kringel, Krishnan, Kr\'olak, Kuehn, Kumar, Kumar, Kumar,
  Kuo, Kutynia, Kwang, Lackey, Lai, Landry, Lang, Lange, Lantz, Lanza, Larson,
  Lartaux-Vollard, Lasky, Laxen, Lazzarini, Lazzaro, Leaci, Leavey, Lee, Lee,
  Lee, Lee, Lee, Lehmann, Lenon, Leon, Leonardi, Leroy, Letendre, Levin, Li,
  Linker, Littenberg, Liu, Liu, Lo, Lockerbie, London, Lord, Lorenzini,
  Loriette, Lormand, Losurdo, Lough, Lousto, Lovelace, L\"uck, Lumaca,
  Lundgren, Lynch, Ma, Macas, Macfoy, Machenschalk, MacInnis, Macleod, Maga\~na
  Hernandez, Maga\~na Sandoval, Maga\~na Zertuche, Magee, Majorana, Maksimovic,
  Man, Mandic, Mangano, Mansell, Manske, Mantovani, Marchesoni, Marion,
  M\'arka, M\'arka, Markakis, Markosyan, Markowitz, Maros, Marquina, Marsh,
  Martelli, Martellini, Martin, Martin, Martynov, Marx, Mason, Massera,
  Masserot, Massinger, Masso-Reid, Mastrogiovanni, Matas, Matichard, Matone,
  Mavalvala, Mazumder, McCarthy, McClelland, McCormick, McCuller, McGuire,
  McIntyre, McIver, McManus, McNeill, McRae, McWilliams, Meacher, Meadors,
  Mehmet, Meidam, Mejuto-Villa, Melatos, Mendell, Mercer, Merilh, Merzougui,
  Meshkov, Messenger, Messick, Metzdorff, Meyers, Miao, Michel, Middleton,
  Mikhailov, Milano, Miller, Miller, Miller, Millhouse, Milovich-Goff,
  Minazzoli, Minenkov, Ming, Mishra, Mitra, Mitrofanov, Mitselmakher,
  Mittleman, Moffa, Moggi, Mogushi, Mohan, Mohapatra, Molina, Montani, Moore,
  Moraru, Moreno, Morisaki, Morriss, Mours, Mow-Lowry, Mueller, Muir,
  Mukherjee, Mukherjee, Mukherjee, Mukund, Mullavey, Munch, Mu\~niz, Muratore,
  Murray, Nagar, Napier, Nardecchia, Naticchioni, Nayak, Neilson, Nelemans,
  Nelson, Nery, Neunzert, Nevin, Newport, Newton, Ng, Nguyen, Nguyen, Nichols,
  Nielsen, Nissanke, Nitz, Noack, Nocera, Nolting, North, Nuttall, Oberling,
  O'Dea, Ogin, Oh, Oh, Ohme, Okada, Oliver, Oppermann, Oram, O'Reilly,
  Ormiston, Ortega, O'Shaughnessy, Ossokine, Ottaway, Overmier, Owen, Pace,
  Page, Page, Pai, Pai, Palamos, Palashov, Palomba, Pal-Singh, Pan, Pan, Pang,
  Pang, Pankow, Pannarale, Pant, Paoletti, Paoli, Papa, Parida, Parker,
  Pascucci, Pasqualetti, Passaquieti, Passuello, Patil, Patricelli, Pearlstone,
  Pedraza, Pedurand, Pekowsky, Pele, Penn, Perez, Perreca, Perri, Pfeiffer,
  Phelps, Piccinni, Pichot, Piergiovanni, Pierro, Pillant, Pinard, Pinto,
  Pirello, Pitkin, Poe, Poggiani, Popolizio, Porter, Post, Powell, Prasad,
  Pratt, Pratten, Predoi, Prestegard, Prijatelj, Principe, Privitera, Prix,
  Prodi, Prokhorov, Puncken, Punturo, Puppo, P\"urrer, Qi, Quetschke, Quintero,
  Quitzow-James, Raab, Rabeling, Radkins, Raffai, Raja, Rajan, Rajbhandari,
  Rakhmanov, Ramirez, Ramos-Buades, Rapagnani, Raymond, Razzano, Read,
  Regimbau, Rei, Reid, Reitze, Ren, Reyes, Ricci, Ricker, Rieger, Riles, Rizzo,
  Robertson, Robie, Robinet, Rocchi, Rolland, Rollins, Roma, Romano, Romano,
  Romel, Romie, Rosi\ifmmode~\acute{n}\else \'{n}\fi{}ska, Ross, Rowan,
  R\"udiger, Ruggi, Rutins, Ryan, Sachdev, Sadecki, Sadeghian, Sakellariadou,
  Salconi, Saleem, Salemi, Samajdar, Sammut, Sampson, Sanchez, Sanchez,
  Sanchis-Gual, Sandberg, Sanders, Sassolas, Sathyaprakash, Saulson, Sauter,
  Savage, Sawadsky, Schale, Scheel, Scheuer, Schmidt, Schmidt, Schnabel,
  Schofield, Sch\"onbeck, Schreiber, Schuette, Schulte, Schutz, Schwalbe,
  Scott, Scott, Seidel, Sellers, Sengupta, Sentenac, Sequino, Sergeev,
  Shaddock, Shaffer, Shah, Shahriar, Shaner, Shao, Shapiro, Shawhan, Sheperd,
  Shoemaker, Shoemaker, Siellez, Siemens, Sieniawska, Sigg, Silva, Singer,
  Singh, Singhal, Sintes, Slagmolen, Smith, Smith, Smith, Somala, Son,
  Sonnenberg, Sorazu, Sorrentino, Souradeep, Spencer, Srivastava, Staats,
  Staley, Steinke, Steinlechner, Steinlechner, Steinmeyer, Stevenson, Stone,
  Stops, Strain, Stratta, Strigin, Strunk, Sturani, Stuver, Summerscales, Sun,
  Sunil, Suresh, Sutton, Swinkels, Szczepa\ifmmode~\acute{n}\else
  \'{n}\fi{}czyk, Tacca, Tait, Talbot, Talukder, Tanner, T\'apai, Taracchini,
  Tasson, Taylor, Taylor, Tewari, Theeg, Thies, Thomas, Thomas, Thomas, Thorne,
  Thorne, Thrane, Tiwari, Tiwari, Tokmakov, Toland, Tonelli, Tornasi,
  Torres-Forn\'e, Torrie, T\"oyr\"a, Travasso, Traylor, Trinastic, Tringali,
  Trozzo, Tsang, Tse, Tso, Tsukada, Tsuna, Tuyenbayev, Ueno, Ugolini,
  Unnikrishnan, Urban, Usman, Vahlbruch, Vajente, Valdes, Vallisneri, van
  Bakel, van Beuzekom, van~den Brand, Van Den~Broeck, Vander-Hyde, van~der
  Schaaf, van Heijningen, van Veggel, Vardaro, Varma, Vass, Vas\'uth, Vecchio,
  Vedovato, Veitch, Veitch, Venkateswara, Venugopalan, Verkindt, Vetrano,
  Vicer\'e, Viets, Vinciguerra, Vine, Vinet, Vitale, Vo, Vocca, Vorvick,
  Vyatchanin, Wade, Wade, Wade, Walet, Walker, Wallace, Walsh, Wang, Wang,
  Wang, Wang, Wang, Ward, Warner, Was, Watchi, Weaver, Wei, Weinert, Weinstein,
  Weiss, Wen, Wessel, We\ss{}els, Westerweck, Westphal, Wette, Whelan,
  Whitcomb, Whiting, Whittle, Wilken, Williams, Williams, Williamson, Willis,
  Willke, Wimmer, Winkler, Wipf, Wittel, Woan, Woehler, Wofford, Wong, Worden,
  Wright, Wu, Wysocki, Xiao, Yamamoto, Yancey, Yang, Yap, Yazback, Yu, Yu,
  Yvert, Zadro\ifmmode~\dot{z}\else \.{z}\fi{}ny, Zanolin, Zelenova, Zendri,
  Zevin, Zhang, Zhang, Zhang, Zhang, Zhao, Zhou, Zhou, Zhu, Zhu, Zimmerman,
  Zucker, \& Zweizig}]{Abbott2017PRL}
Abbott, B.~P., Abbott, R., Abbott, T.~D., {et~al.} 2017, PhRvL, 119, 161101,
  \dodoi{10.1103/PhysRevLett.119.161101}

\bibitem[{Abbott {et~al.}(2018)Abbott, Abbott, Abbott, Acernese, Ackley, Adams,
  Adams, Addesso, Adhikari, Adya, Affeldt, Agarwal, Agathos, Agatsuma,
  Aggarwal, Aguiar, Aiello, Ain, Ajith, Allen, Allen, Allocca, Aloy, Altin,
  Amato, Ananyeva, Anderson, Anderson, Angelova, Antier, Appert, Arai, Araya,
  Areeda, Ar\`ene, Arnaud, Arun, Ascenzi, Ashton, Ast, Aston, Astone, Atallah,
  Aubin, Aufmuth, Aulbert, AultONeal, Austin, Avila-Alvarez, Babak, Bacon,
  Badaracco, Bader, Bae, Baker, Baldaccini, Ballardin, Ballmer, Banagiri,
  Barayoga, Barclay, Barish, Barker, Barkett, Barnum, Barone, Barr, Barsotti,
  Barsuglia, Barta, Bartlett, Bartos, Bassiri, Basti, Batch, Bawaj, Bayley,
  Bazzan, B\'ecsy, Beer, Bejger, Belahcene, Bell, Beniwal, Bensch, Berger,
  Bergmann, Bernuzzi, Bero, Berry, Bersanetti, Bertolini, Betzwieser, Bhandare,
  Bilenko, Bilgili, Billingsley, Billman, Birch, Birney, Birnholtz, Biscans,
  Biscoveanu, Bisht, Bitossi, Bizouard, Blackburn, Blackman, Blair, Blair,
  Blair, Bloemen, Bock, Bode, Boer, Boetzel, Bogaert, Bohe, Bondu, Bonilla,
  Bonnand, Booker, Boom, Booth, Bork, Boschi, Bose, Bossie, Bossilkov, Bosveld,
  Bouffanais, Bozzi, Bradaschia, Brady, Bramley, Branchesi, Brau, Briant,
  Brighenti, Brillet, Brinkmann, Brisson, Brockill, Brooks, Brown, Brunett,
  Buchanan, Buikema, Bulik, Bulten, Buonanno, Buskulic, Buy, Byer, Cabero,
  Cadonati, Cagnoli, Cahillane, Calder\'on~Bustillo, Callister, Calloni, Camp,
  Canepa, Canizares, Cannon, Cao, Cao, Capano, Capocasa, Carbognani, Caride,
  Carney, Carullo, Casanueva~Diaz, Casentini, Caudill, Cavagli\`a, Cavalier,
  Cavalieri, Cella, Cepeda, Cerd\'a-Dur\'an, Cerretani, Cesarini, Chaibi,
  Chamberlin, Chan, Chao, Charlton, Chase, Chassande-Mottin, Chatterjee,
  Chatziioannou, Cheeseboro, Chen, Chen, Chen, Cheng, Chia, Chincarini,
  Chiummo, Chmiel, Cho, Cho, Chow, Christensen, Chu, Chua, Chua, Chung, Chung,
  Ciani, Ciobanu, Ciolfi, Cipriano, Cirelli, Cirone, Clara, Clark, Clearwater,
  Cleva, Cocchieri, Coccia, Cohadon, Cohen, Colla, Collette, Collins, Cominsky,
  Constancio, Conti, Cooper, Corban, Corbitt, Cordero-Carri\'on, Corley,
  Cornish, Corsi, Cortese, Costa, Cotesta, Coughlin, Coughlin, Coulon,
  Countryman, Couvares, Covas, Cowan, Coward, Cowart, Coyne, Coyne, Creighton,
  Creighton, Cripe, Crowder, Cullen, Cumming, Cunningham, Cuoco, Canton,
  D\'alya, Danilishin, D'Antonio, Danzmann, Dasgupta, Da~Silva~Costa, Dattilo,
  Dave, Davier, Davis, Daw, Day, DeBra, Deenadayalan, Degallaix, De~Laurentis,
  Del\'eglise, Del~Pozzo, Demos, Denker, Dent, De~Pietri, Derby, Dergachev,
  De~Rosa, De~Rossi, DeSalvo, de~Varona, Dhurandhar, D\'{\i}az, Dietrich,
  Di~Fiore, Di~Giovanni, Di~Girolamo, Di~Lieto, Ding, Di~Pace, Di~Palma,
  Di~Renzo, Dmitriev, Doctor, Dolique, Donovan, Dooley, Doravari, Dorrington,
  Dovale~\'Alvarez, Downes, Drago, Dreissigacker, Driggers, Du, Dupej, Dwyer,
  Easter, Edo, Edwards, Effler, Eggenstein, Ehrens, Eichholz, Eikenberry,
  Eisenmann, Eisenstein, Essick, Estelles, Estevez, Etienne, Etzel, Evans,
  Evans, Fafone, Fair, Fairhurst, Fan, Farinon, Farr, Farr, Fauchon-Jones,
  Favata, Fays, Fee, Fehrmann, Feicht, Fejer, Feng, Fernandez-Galiana,
  Ferrante, Ferreira, Ferrini, Fidecaro, Fiori, Fiorucci, Fishbach, Fisher,
  Fishner, Fitz-Axen, Flaminio, Fletcher, Fong, Font, Forsyth, Forsyth,
  Fournier, Frasca, Frasconi, Frei, Freise, Frey, Frey, Fritschel, Frolov,
  Fulda, Fyffe, Gabbard, Gadre, Gaebel, Gair, Gammaitoni, Ganija, Gaonkar,
  Garcia, Garc\'{\i}a-Quir\'os, Garufi, Gateley, Gaudio, Gaur, Gayathri, Gemme,
  Genin, Gennai, George, George, Gergely, Germain, Ghonge, Ghosh, Ghosh, Ghosh,
  Giacomazzo, Giaime, Giardina, Giazotto, Gill, Giordano, Glover, Goetz, Goetz,
  Goncharov, Gonz\'alez, Gonzalez~Castro, Gopakumar, Gorodetsky, Gossan,
  Gosselin, Gouaty, Grado, Graef, Granata, Grant, Gras, Gray, Greco, Green,
  Green, Gretarsson, Groot, Grote, Grunewald, Gruning, Guidi, Gulati, Guo,
  Gupta, Gupta, Gushwa, Gustafson, Gustafson, Halim, Hall, Hall, Hamilton,
  Hamilton, Hammond, Haney, Hanke, Hanks, Hanna, Hannam, Hannuksela, Hanson,
  Hardwick, Harms, Harry, Harry, Hart, Haster, Haughian, Healy, Heidmann,
  Heintze, Heitmann, Hello, Hemming, Hendry, Heng, Hennig, Heptonstall,
  Hernandez, Heurs, Hild, Hinderer, Ho, Hoak, Hochheim, Hofman, Holland, Holt,
  Holz, Hopkins, Horst, Hough, Houston, Howell, Hreibi, Huerta, Huet, Hughey,
  Hulko, Husa, Huttner, Huynh-Dinh, Iess, Indik, Ingram, Inta, Intini, Irwin,
  Isa, Isac, Isi, Iyer, Izumi, Jacqmin, Jani, Jaranowski, Johnson, Johnson,
  Jones, Jones, Jonker, Ju, Junker, Kalaghatgi, Kalogera, Kamai, Kandhasamy,
  Kang, Kanner, Kapadia, Karki, Karvinen, Kasprzack, Katolik, Katsanevas,
  Katsavounidis, Katzman, Kaufer, Kawabe, Keerthana, K\'ef\'elian, Keitel,
  Kemball, Kennedy, Key, Khalili, Khamesra, Khan, Khan, Khan, Khan, Khazanov,
  Kijbunchoo, Kim, Kim, Kim, Kim, Kim, Kim, King, King, Kinley-Hanlon,
  Kirchhoff, Kissel, Kleybolte, Klimenko, Knowles, Koch, Koehlenbeck, Koley,
  Kondrashov, Kontos, Korobko, Korth, Kowalska, Kozak, Kr\"amer, Kringel,
  Krishnan, Kr\'olak, Kuehn, Kumar, Kumar, Kumar, Kuo, Kutynia, Kwang, Lackey,
  Lai, Landry, Landry, Lang, Lange, Lantz, Lanza, Lartaux-Vollard, Lasky,
  Laxen, Lazzarini, Lazzaro, Leaci, Leavey, Lee, Lee, Lee, Lee, Lee, Lehmann,
  Lenon, Leonardi, Leroy, Letendre, Levin, Li, Li, Li, Linker, Littenberg, Liu,
  Liu, Lo, Lockerbie, London, Longo, Lorenzini, Loriette, Lormand, Losurdo,
  Lough, Lousto, Lovelace, L\"uck, Lumaca, Lundgren, Lynch, Ma, Macas, Macfoy,
  Machenschalk, MacInnis, Macleod, Maga\~na Hernandez, Maga\~na Sandoval,
  Maga\~na Zertuche, Magee, Majorana, Maksimovic, Man, Mandic, Mangano,
  Mansell, Manske, Mantovani, Marchesoni, Marion, M\'arka, M\'arka, Markakis,
  Markosyan, Markowitz, Maros, Marquina, Martelli, Martellini, Martin, Martin,
  Martynov, Mason, Massera, Masserot, Massinger, Masso-Reid, Mastrogiovanni,
  Matas, Matichard, Matone, Mavalvala, Mazumder, McCann, McCarthy, McClelland,
  McCormick, McCuller, McGuire, McIver, McManus, McRae, McWilliams, Meacher,
  Meadors, Mehmet, Meidam, Mejuto-Villa, Melatos, Mendell, Mendoza-Gandara,
  Mercer, Mereni, Merilh, Merzougui, Meshkov, Messenger, Messick, Metzdorff,
  Meyers, Miao, Michel, Middleton, Mikhailov, Milano, Miller, Miller, Miller,
  Miller, Millhouse, Mills, Milovich-Goff, Minazzoli, Minenkov, Ming, Mishra,
  Mitra, Mitrofanov, Mitselmakher, Mittleman, Moffa, Mogushi, Mohan, Mohapatra,
  Montani, Moore, Moraru, Moreno, Morisaki, Mours, Mow-Lowry, Mueller, Muir,
  Mukherjee, Mukherjee, Mukherjee, Mukund, Mullavey, Munch, Mu\~niz, Muratore,
  Murray, Nagar, Napier, Nardecchia, Naticchioni, Nayak, Neilson, Nelemans,
  Nelson, Nery, Neunzert, Nevin, Newport, Ng, Ng, Nguyen, Nguyen, Nichols,
  Nielsen, Nissanke, Nitz, Nocera, Nolting, North, Nuttall, Obergaulinger,
  Oberling, O'Brien, O'Dea, Ogin, Oh, Oh, Ohme, Ohta, Okada, Oliver, Oppermann,
  Oram, O'Reilly, Ormiston, Ortega, O'Shaughnessy, Ossokine, Ottaway, Overmier,
  Owen, Pace, Pagano, Page, Page, Pai, Pai, Palamos, Palashov, Palomba,
  Pal-Singh, Pan, Pan, Pang, Pang, Pankow, Pannarale, Pant, Paoletti, Paoli,
  Papa, Parida, Parker, Pascucci, Pasqualetti, Passaquieti, Passuello, Patil,
  Patricelli, Pearlstone, Pedersen, Pedraza, Pedurand, Pekowsky, Pele, Penn,
  Perego, Perez, Perreca, Perri, Pfeiffer, Phelps, Phukon, Piccinni, Pichot,
  Piergiovanni, Pierro, Pillant, Pinard, Pinto, Pirello, Pitkin, Poggiani,
  Popolizio, Porter, Possenti, Post, Powell, Prasad, Pratt, Pratten, Predoi,
  Prestegard, Principe, Privitera, Prodi, Prokhorov, Puncken, Punturo, Puppo,
  P\"urrer, Qi, Quetschke, Quintero, Quitzow-James, Raab, Rabeling, Radkins,
  Raffai, Raja, Rajan, Rajbhandari, Rakhmanov, Ramirez, Ramos-Buades, Rana,
  Rapagnani, Raymond, Razzano, Read, Regimbau, Rei, Reid, Reitze, Ren, Ricci,
  Ricker, Riemenschneider, Riles, Rizzo, Robertson, Robie, Robinet, Robson,
  Rocchi, Rolland, Rollins, Roma, Romano, Romel, Romie,
  Rosi\ifmmode~\acute{n}\else \'{n}\fi{}ska, Ross, Rowan, R\"udiger, Ruggi,
  Rutins, Ryan, Sachdev, Sadecki, Sakellariadou, Salconi, Saleem, Salemi,
  Samajdar, Sammut, Sampson, Sanchez, Sanchez, Sanchis-Gual, Sandberg, Sanders,
  Sarin, Sassolas, Sathyaprakash, Saulson, Sauter, Savage, Sawadsky, Schale,
  Scheel, Scheuer, Schmidt, Schnabel, Schofield, Sch\"onbeck, Schreiber,
  Schuette, Schulte, Schutz, Schwalbe, Scott, Scott, Seidel, Sellers, Sengupta,
  Sentenac, Sequino, Sergeev, Setyawati, Shaddock, Shaffer, Shah, Shahriar,
  Shaner, Shao, Shapiro, Shawhan, Shen, Shoemaker, Shoemaker, Siellez, Siemens,
  Sieniawska, Sigg, Silva, Singer, Singh, Singhal, Sintes, Slagmolen,
  Slaven-Blair, Smith, Smith, Smith, Somala, Son, Sorazu, Sorrentino,
  Souradeep, Spencer, Srivastava, Staats, Steinke, Steinlechner, Steinlechner,
  Steinmeyer, Steltner, Stevenson, Stocks, Stone, Stops, Strain, Stratta,
  Strigin, Strunk, Sturani, Stuver, Summerscales, Sun, Sunil, Suresh, Sutton,
  Swinkels, Szczepa\ifmmode~\acute{n}\else \'{n}\fi{}czyk, Tacca, Tait, Talbot,
  Talukder, Tanner, T\'apai, Taracchini, Tasson, Taylor, Taylor, Tewari, Theeg,
  Thies, Thomas, Thomas, Thomas, Thorne, Thrane, Tiwari, Tiwari, Tokmakov,
  Toland, Tonelli, Tornasi, Torres-Forn\'e, Torrie, T\"oyr\"a, Travasso,
  Traylor, Trinastic, Tringali, Trovato, Trozzo, Tsang, Tse, Tso, Tsuna,
  Tsukada, Tuyenbayev, Ueno, Ugolini, Urban, Usman, Vahlbruch, Vajente, Valdes,
  van Bakel, van Beuzekom, van~den Brand, Van Den~Broeck, Vander-Hyde, van~der
  Schaaf, van Heijningen, van Veggel, Vardaro, Varma, Vass, Vas\'uth, Vecchio,
  Vedovato, Veitch, Veitch, Venkateswara, Venugopalan, Verkindt, Vetrano,
  Vicer\'e, Viets, Vinciguerra, Vine, Vinet, Vitale, Vo, Vocca, Vorvick,
  Vyatchanin, Wade, Wade, Wade, Walet, Walker, Wallace, Walsh, Wang, Wang,
  Wang, Wang, Wang, Ward, Warner, Was, Watchi, Weaver, Wei, Weinert, Weinstein,
  Weiss, Wellmann, Wen, Wessel, We\ss{}els, Westerweck, Wette, Whelan, Whiting,
  Whittle, Wilken, Williams, Williams, Williamson, Willis, Willke, Wimmer,
  Winkler, Wipf, Wittel, Woan, Woehler, Wofford, Wong, Worden, Wright, Wu,
  Wysocki, Xiao, Yam, Yamamoto, Yancey, Yang, Yap, Yazback, Yu, Yu, Yvert,
  Zadro\ifmmode~\dot{z}\else \.{z}\fi{}ny, Zanolin, Zelenova, Zendri, Zevin,
  Zhang, Zhang, Zhang, Zhang, Zhang, Zhao, Zhou, Zhou, Zhu, Zhu, Zimmerman,
  Zlochower, Zucker, \& Zweizig}]{Abbott2018PRL}
---. 2018, PhRvL, 121, 161101, \dodoi{10.1103/PhysRevLett.121.161101}

\bibitem[{Abbott {et~al.}(2019)Abbott, Abbott, Abbott, Acernese, Ackley, Adams,
  Adams, Addesso, Adhikari, Adya, Affeldt, Agarwal, Agathos, Agatsuma,
  Aggarwal, Aguiar, Aiello, Ain, Ajith, Allen, Allen, Allocca, Aloy, Altin,
  Amato, Ananyeva, Anderson, Anderson, Angelova, Antier, Appert, Arai, Araya,
  Areeda, Ar\`ene, Arnaud, Arun, Ascenzi, Ashton, Ast, Aston, Astone, Atallah,
  Aubin, Aufmuth, Aulbert, AultONeal, Austin, Avila-Alvarez, Babak, Bacon,
  Badaracco, Bader, Bae, Baker, Baldaccini, Ballardin, Ballmer, Banagiri,
  Barayoga, Barclay, Barish, Barker, Barkett, Barnum, Barone, Barr, Barsotti,
  Barsuglia, Barta, Bartlett, Bartos, Bassiri, Basti, Batch, Bawaj, Bayley,
  Bazzan, B\'ecsy, Beer, Bejger, Belahcene, Bell, Beniwal, Bensch, Berger,
  Bergmann, Bernuzzi, Bero, Berry, Bersanetti, Bertolini, Betzwieser, Bhandare,
  Bilenko, Bilgili, Billingsley, Billman, Birch, Birney, Birnholtz, Biscans,
  Biscoveanu, Bisht, Bitossi, Bizouard, Blackburn, Blackman, Blair, Blair,
  Blair, Bloemen, Bock, Bode, Boer, Boetzel, Bogaert, Bohe, Bondu, Bonilla,
  Bonnand, Booker, Boom, Booth, Bork, Boschi, Bose, Bossie, Bossilkov, Bosveld,
  Bouffanais, Bozzi, Bradaschia, Brady, Bramley, Branchesi, Brau, Briant,
  Brighenti, Brillet, Brinkmann, Brisson, Brockill, Brooks, Brown, Brunett,
  Buchanan, Buikema, Bulik, Bulten, Buonanno, Buskulic, Buy, Byer, Cabero,
  Cadonati, Cagnoli, Cahillane, Bustillo, Callister, Calloni, Camp, Canepa,
  Canizares, Cannon, Cao, Cao, Capano, Capocasa, Carbognani, Caride, Carney,
  Carullo, Diaz, Casentini, Caudill, Cavagli\`a, Cavalier, Cavalieri, Cella,
  Cepeda, Cerd\'a-Dur\'an, Cerretani, Cesarini, Chaibi, Chamberlin, Chan, Chao,
  Charlton, Chase, Chassande-Mottin, Chatterjee, Chatziioannou, Cheeseboro,
  Chen, Chen, Chen, Cheng, Chia, Chincarini, Chiummo, Chmiel, Cho, Cho, Chow,
  Christensen, Chu, Chua, Chua, Chung, Chung, Ciani, Ciobanu, Ciolfi, Cipriano,
  Cirelli, Cirone, Clara, Clark, Clearwater, Cleva, Cocchieri, Coccia, Cohadon,
  Cohen, Colla, Collette, Collins, Cominsky, Constancio, Conti, Cooper, Corban,
  Corbitt, Cordero-Carri\'on, Corley, Cornish, Corsi, Cortese, Costa, Cotesta,
  Coughlin, Coughlin, Coulon, Countryman, Couvares, Covas, Cowan, Coward,
  Cowart, Coyne, Coyne, Creighton, Creighton, Cripe, Crowder, Cullen, Cumming,
  Cunningham, Cuoco, Canton, D\'alya, Danilishin, D'Antonio, Danzmann,
  Dasgupta, Costa, Dattilo, Dave, Davier, Davis, Daw, Day, DeBra, Deenadayalan,
  Degallaix, De~Laurentis, Del\'eglise, Del~Pozzo, Demos, Denker, Dent,
  De~Pietri, Derby, Dergachev, De~Rosa, De~Rossi, DeSalvo, de~Varona,
  Dhurandhar, D\'{\i}az, Dietrich, Di~Fiore, Di~Giovanni, Di~Girolamo,
  Di~Lieto, Ding, Di~Pace, Di~Palma, Di~Renzo, Dmitriev, Doctor, Dolique,
  Donovan, Dooley, Doravari, Dorrington, \'Alvarez, Downes, Drago,
  Dreissigacker, Driggers, Du, Dudi, Dupej, Dwyer, Easter, Edo, Edwards,
  Effler, Eggenstein, Ehrens, Eichholz, Eikenberry, Eisenmann, Eisenstein,
  Essick, Estelles, Estevez, Etienne, Etzel, Evans, Evans, Fafone, Fair,
  Fairhurst, Fan, Farinon, Farr, Farr, Fauchon-Jones, Favata, Fays, Fee,
  Fehrmann, Feicht, Fejer, Feng, Fernandez-Galiana, Ferrante, Ferreira,
  Ferrini, Fidecaro, Fiori, Fiorucci, Fishbach, Fisher, Fishner, Fitz-Axen,
  Flaminio, Fletcher, Fong, Font, Forsyth, Forsyth, Fournier, Frasca, Frasconi,
  Frei, Freise, Frey, Frey, Fritschel, Frolov, Fulda, Fyffe, Gabbard, Gadre,
  Gaebel, Gair, Gammaitoni, Ganija, Gaonkar, Garcia, Garc\'{\i}a-Quir\'os,
  Garufi, Gateley, Gaudio, Gaur, Gayathri, Gemme, Genin, Gennai, George,
  George, Gergely, Germain, Ghonge, Ghosh, Ghosh, Ghosh, Giacomazzo, Giaime,
  Giardina, Giazotto, Gill, Giordano, Glover, Goetz, Goetz, Goncharov,
  Gonz\'alez, Castro, Gopakumar, Gorodetsky, Gossan, Gosselin, Gouaty, Grado,
  Graef, Granata, Grant, Gras, Gray, Greco, Green, Green, Gretarsson, Groot,
  Grote, Grunewald, Gruning, Guidi, Gulati, Guo, Gupta, Gupta, Gushwa,
  Gustafson, Gustafson, Halim, Hall, Hall, Hamilton, Hamilton, Hammond, Haney,
  Hanke, Hanks, Hanna, Hannam, Hannuksela, Hanson, Hardwick, Harms, Harry,
  Harry, Hart, Haster, Haughian, Healy, Heidmann, Heintze, Heitmann, Hello,
  Hemming, Hendry, Heng, Hennig, Heptonstall, Hernandez, Heurs, Hild, Hinderer,
  Hoak, Hochheim, Hofman, Holland, Holt, Holz, Hopkins, Horst, Hough, Houston,
  Howell, Hreibi, Huerta, Huet, Hughey, Hulko, Husa, Huttner, Huynh-Dinh, Iess,
  Indik, Ingram, Inta, Intini, Isa, Isac, Isi, Iyer, Izumi, Jacqmin, Jani,
  Jaranowski, Johnson, Johnson, Jones, Jones, Jonker, Ju, Junker, Kalaghatgi,
  Kalogera, Kamai, Kandhasamy, Kang, Kanner, Kapadia, Karki, Karvinen,
  Kasprzack, Kastaun, Katolik, Katsanevas, Katsavounidis, Katzman, Kaufer,
  Kawabe, Keerthana, K\'ef\'elian, Keitel, Kemball, Kennedy, Key, Khalili,
  Khamesra, Khan, Khan, Khan, Khan, Khazanov, Kijbunchoo, Kim, Kim, Kim, Kim,
  Kim, Kim, King, King, Kinley-Hanlon, Kirchhoff, Kissel, Kleybolte, Klimenko,
  Knowles, Koch, Koehlenbeck, Koley, Kondrashov, Kontos, Korobko, Korth,
  Kowalska, Kozak, Kr\"amer, Kringel, Krishnan, Kr\'olak, Kuehn, Kumar, Kumar,
  Kumar, Kuo, Kutynia, Kwang, Lackey, Lai, Landry, Landry, Lang, Lange, Lantz,
  Lanza, Lartaux-Vollard, Lasky, Laxen, Lazzarini, Lazzaro, Leaci, Leavey, Lee,
  Lee, Lee, Lee, Lee, Lehmann, Lenon, Leonardi, Leroy, Letendre, Levin, Li, Li,
  Li, Linker, Littenberg, Liu, Liu, Lo, Lockerbie, London, Longo, Lorenzini,
  Loriette, Lormand, Losurdo, Lough, Lousto, Lovelace, L\"uck, Lumaca,
  Lundgren, Lynch, Ma, Macas, Macfoy, Machenschalk, MacInnis, Macleod,
  Hernandez, Maga\~na Sandoval, Zertuche, Magee, Majorana, Maksimovic, Man,
  Mandic, Mangano, Mansell, Manske, Mantovani, Marchesoni, Marion, M\'arka,
  M\'arka, Markakis, Markosyan, Markowitz, Maros, Marquina, Martelli,
  Martellini, Martin, Martin, Martynov, Mason, Massera, Masserot, Massinger,
  Masso-Reid, Mastrogiovanni, Matas, Matichard, Matone, Mavalvala, Mazumder,
  McCann, McCarthy, McClelland, McCormick, McCuller, McGuire, McIver, McManus,
  McRae, McWilliams, Meacher, Meadors, Mehmet, Meidam, Mejuto-Villa, Melatos,
  Mendell, Mendoza-Gandara, Mercer, Mereni, Merilh, Merzougui, Meshkov,
  Messenger, Messick, Metzdorff, Meyers, Miao, Michel, Middleton, Mikhailov,
  Milano, Miller, Miller, Miller, Miller, Millhouse, Mills, Milovich-Goff,
  Minazzoli, Minenkov, Ming, Mishra, Mitra, Mitrofanov, Mitselmakher,
  Mittleman, Moffa, Mogushi, Mohan, Mohapatra, Montani, Moore, Moraru, Moreno,
  Morisaki, Mours, Mow-Lowry, Mueller, Muir, Mukherjee, Mukherjee, Mukherjee,
  Mukund, Mullavey, Munch, Mu\~niz, Muratore, Murray, Nagar, Napier,
  Nardecchia, Naticchioni, Nayak, Neilson, Nelemans, Nelson, Nery, Neunzert,
  Nevin, Newport, Ng, Ng, Nguyen, Nguyen, Nichols, Nielsen, Nissanke, Nitz,
  Nocera, Nolting, North, Nuttall, Obergaulinger, Oberling, O'Brien, O'Dea,
  Ogin, Oh, Oh, Ohme, Ohta, Okada, Oliver, Oppermann, Oram, O'Reilly, Ormiston,
  Ortega, O'Shaughnessy, Ossokine, Ottaway, Overmier, Owen, Pace, Pagano, Page,
  Page, Pai, Pai, Palamos, Palashov, Palomba, Pal-Singh, Pan, Pan, Pang, Pang,
  Pankow, Pannarale, Pant, Paoletti, Paoli, Papa, Parida, Parker, Pascucci,
  Pasqualetti, Passaquieti, Passuello, Patil, Patricelli, Pearlstone, Pedersen,
  Pedraza, Pedurand, Pekowsky, Pele, Penn, Perez, Perreca, Perri, Pfeiffer,
  Phelps, Phukon, Piccinni, Pichot, Piergiovanni, Pierro, Pillant, Pinard,
  Pinto, Pirello, Pitkin, Poggiani, Popolizio, Porter, Possenti, Post, Powell,
  Prasad, Pratt, Pratten, Predoi, Prestegard, Principe, Privitera, Prodi,
  Prokhorov, Puncken, Punturo, Puppo, P\"urrer, Qi, Quetschke, Quintero,
  Quitzow-James, Raab, Rabeling, Radkins, Raffai, Raja, Rajan, Rajbhandari,
  Rakhmanov, Ramirez, Ramos-Buades, Rana, Rapagnani, Raymond, Razzano, Read,
  Regimbau, Rei, Reid, Reitze, Ren, Ricci, Ricker, Riemenschneider, Riles,
  Rizzo, Robertson, Robie, Robinet, Robson, Rocchi, Rolland, Rollins, Roma,
  Romano, Romel, Romie, Rosi\ifmmode~\acute{n}\else \'{n}\fi{}ska, Ross, Rowan,
  R\"udiger, Ruggi, Rutins, Ryan, Sachdev, Sadecki, Sakellariadou, Salconi,
  Saleem, Salemi, Samajdar, Sammut, Sampson, Sanchez, Sanchez, Sanchis-Gual,
  Sandberg, Sanders, Sarin, Sassolas, Sathyaprakash, Saulson, Sauter, Savage,
  Sawadsky, Schale, Scheel, Scheuer, Schmidt, Schnabel, Schofield, Sch\"onbeck,
  Schreiber, Schuette, Schulte, Schutz, Schwalbe, Scott, Scott, Seidel,
  Sellers, Sengupta, Sentenac, Sequino, Sergeev, Setyawati, Shaddock, Shaffer,
  Shah, Shahriar, Shaner, Shao, Shapiro, Shawhan, Shen, Shoemaker, Shoemaker,
  Siellez, Siemens, Sieniawska, Sigg, Silva, Singer, Singh, Singhal, Sintes,
  Slagmolen, Slaven-Blair, Smith, Smith, Smith, Somala, Son, Sorazu,
  Sorrentino, Souradeep, Spencer, Srivastava, Staats, Steinke, Steinlechner,
  Steinlechner, Steinmeyer, Steltner, Stevenson, Stocks, Stone, Stops, Strain,
  Stratta, Strigin, Strunk, Sturani, Stuver, Summerscales, Sun, Sunil, Suresh,
  Sutton, Swinkels, Szczepa\ifmmode~\acute{n}\else \'{n}\fi{}czyk, Tacca, Tait,
  Talbot, Talukder, Tanner, T\'apai, Taracchini, Tasson, Taylor, Taylor,
  Tewari, Theeg, Thies, Thomas, Thomas, Thomas, Thorne, Thrane, Tiwari, Tiwari,
  Tokmakov, Toland, Tonelli, Tornasi, Torres-Forn\'e, Torrie, T\"oyr\"a,
  Travasso, Traylor, Trinastic, Tringali, Trozzo, Tsang, Tse, Tso, Tsuna,
  Tsukada, Tuyenbayev, Ueno, Ugolini, Urban, Usman, Vahlbruch, Vajente, Valdes,
  van Bakel, van Beuzekom, van~den Brand, Van Den~Broeck, Vander-Hyde, van~der
  Schaaf, van Heijningen, van Veggel, Vardaro, Varma, Vass, Vas\'uth, Vecchio,
  Vedovato, Veitch, Veitch, Venkateswara, Venugopalan, Verkindt, Vetrano,
  Vicer\'e, Viets, Vinciguerra, Vine, Vinet, Vitale, Vo, Vocca, Vorvick,
  Vyatchanin, Wade, Wade, Wade, Walet, Walker, Wallace, Walsh, Wang, Wang,
  Wang, Wang, Wang, Ward, Warner, Was, Watchi, Weaver, Wei, Weinert, Weinstein,
  Weiss, Wellmann, Wen, Wessel, We\ss{}els, Westerweck, Wette, Whelan, Whiting,
  Whittle, Wilken, Williams, Williams, Williamson, Willis, Willke, Wimmer,
  Winkler, Wipf, Wittel, Woan, Woehler, Wofford, Wong, Worden, Wright, Wu,
  Wysocki, Xiao, Yam, Yamamoto, Yancey, Yang, Yap, Yazback, Yu, Yu, Yvert,
  Zadro\ifmmode~\dot{z}\else \.{z}\fi{}ny, Zanolin, Zelenova, Zendri, Zevin,
  Zhang, Zhang, Zhang, Zhang, Zhang, Zhao, Zhou, Zhou, Zhu, Zhu, Zimmerman,
  Zlochower, Zucker, \& Zweizig}]{Abbott2019PRX}
---. 2019, PhRvX, 9, 011001, \dodoi{10.1103/PhysRevX.9.011001}

\bibitem[{Abbott {et~al.}(2020)Abbott, Abbott, Abraham, Acernese, Ackley,
  Adams, Adhikari, Adya, Affeldt, Agathos, Agatsuma, Aggarwal, Aguiar, Aich,
  Aiello, Ain, Ajith, Akcay, Allen, Allocca, Altin, Amato, Anand, Ananyeva,
  Anderson, Anderson, Angelova, Ansoldi, Antier, Appert, Arai, Araya, Areeda,
  Ar{\`{e}}ne, Arnaud, Aronson, Arun, Asali, Ascenzi, Ashton, Aston, Astone,
  Aubin, Aufmuth, AultONeal, Austin, Avendano, Babak, Bacon, Badaracco, Bader,
  Bae, Baer, Baird, Baldaccini, Ballardin, Ballmer, Bals, Balsamo, Baltus,
  Banagiri, Bankar, Bankar, Barayoga, Barbieri, Barish, Barker, Barkett,
  Barneo, Barone, Barr, Barsotti, Barsuglia, Barta, Bartlett, Bartos, Bassiri,
  Basti, Bawaj, Bayley, Bazzan, B{\'{e}}csy, Bejger, Belahcene, Bell, Beniwal,
  Benjamin, Benkel, Bentley, Bergamin, Berger, Bergmann, Bernuzzi, Berry,
  Bersanetti, Bertolini, Betzwieser, Bhandare, Bhandari, Bidler, Biggs,
  Bilenko, Billingsley, Birney, Birnholtz, Biscans, Bischi, Biscoveanu, Bisht,
  Bissenbayeva, Bitossi, Bizouard, Blackburn, Blackman, Blair, Blair, Blair,
  Bobba, Bode, Boer, Boetzel, Bogaert, Bondu, Bonilla, Bonnand, Booker, Boom,
  Bork, Boschi, Bose, Bossilkov, Bosveld, Bouffanais, Bozzi, Bradaschia, Brady,
  Bramley, Branchesi, Brau, Breschi, Briant, Briggs, Brighenti, Brillet,
  Brinkmann, Brito, Brockill, Brooks, Brooks, Brown, Brunett, Bruno, Bruntz,
  Buikema, Bulik, Bulten, Buonanno, Buskulic, Byer, Cabero, Cadonati, Cagnoli,
  Cahillane, Bustillo, Callaghan, Callister, Calloni, Camp, Canepa, Cannon,
  Cao, Cao, Carapella, Carbognani, Caride, Carney, Carullo, Diaz, Casentini,
  Casta{\~{n}}eda, Caudill, Cavagli{\`{a}}, Cavalier, Cavalieri, Cella,
  Cerd{\'{a}}-Dur{\'{a}}n, Cesarini, Chaibi, Chakravarti, Chan, Chan, Chao,
  Charlton, Chase, Chassande-Mottin, Chatterjee, Chaturvedi, Chatziioannou,
  Chen, Chen, Chen, Cheng, Cheong, Chia, Chiadini, Chierici, Chincarini,
  Chiummo, Cho, Cho, Cho, Christensen, Chu, Chua, Chung, Chung, Ciani,
  Ciecielag, Cie{\'{s}}lar, Ciobanu, Ciolfi, Cipriano, Cirone, Clara, Clark,
  Clearwater, Clesse, Cleva, Coccia, Cohadon, Cohen, Colleoni, Collette,
  Collins, Colpi, Constancio, Conti, Cooper, Corban, Corbitt,
  Cordero-Carri{\'{o}}n, Corezzi, Corley, Cornish, Corre, Corsi, Cortese,
  Costa, Cotesta, Coughlin, Coughlin, Coulon, Countryman, Couvares, Covas,
  Coward, Cowart, Coyne, Coyne, Creighton, Creighton, Cripe, Croquette,
  Crowder, Cudell, Cullen, Cumming, Cummings, Cunningham, Cuoco, Curylo,
  Canton, D{\'{a}}lya, Dana, Daneshgaran-Bajastani, D'Angelo, Danilishin,
  D'Antonio, Danzmann, Darsow-Fromm, Dasgupta, Datrier, Dattilo, Dave, Davier,
  Davies, Davis, Daw, DeBra, Deenadayalan, Degallaix, Laurentis,
  Del{\'{e}}glise, Delfavero, Lillo, Pozzo, DeMarchi, D'Emilio, Demos, Dent,
  Pietri, Rosa, Rossi, DeSalvo, de~Varona, Dhurandhar, D{\'{\i}}az, Diaz-Ortiz,
  Dietrich, Fiore, Fronzo, Giorgio, Giovanni, Giovanni, Girolamo, Lieto, Ding,
  Pace, Palma, Renzo, Divakarla, Dmitriev, Doctor, Donovan, Dooley, Doravari,
  Dorrington, Downes, Drago, Driggers, Du, Ducoin, Dupej, Durante, D'Urso,
  Dwyer, Easter, Eddolls, Edelman, Edo, Edy, Effler, Ehrens, Eichholz,
  Eikenberry, Eisenmann, Eisenstein, Ejlli, Errico, Essick, Estelles, Estevez,
  Etienne, Etzel, Evans, Evans, Ewing, Fafone, Fairhurst, Fan, Farinon, Farr,
  Farr, Fauchon-Jones, Favata, Fays, Fazio, Feicht, Fejer, Feng, Fenyvesi,
  Ferguson, Fernandez-Galiana, Ferrante, Ferreira, Ferreira, Fidecaro, Fiori,
  Fiorucci, Fishbach, Fisher, Fittipaldi, Fitz-Axen, Fiumara, Flaminio, Floden,
  Flynn, Fong, Font, Forsyth, Fournier, Frasca, Frasconi, Frei, Freise, Frey,
  Frey, Fritschel, Frolov, Fronz{\`{e}}, Fulda, Fyffe, Gabbard, Gadre, Gaebel,
  Gair, Galaudage, Ganapathy, Ganguly, Gaonkar, Garc{\'{\i}}a-Quir{\'{o}}s,
  Garufi, Gateley, Gaudio, Gayathri, Gemme, Genin, Gennai, George, George,
  Gergely, Ghonge, Ghosh, Ghosh, Ghosh, Giacomazzo, Giaime, Giardina, Gibson,
  Gier, Gill, Glanzer, Gniesmer, Godwin, Goetz, Goetz, Gohlke, Goncharov,
  Gonz{\'{a}}lez, Gopakumar, Gossan, Gosselin, Gouaty, Grace, Grado, Granata,
  Grant, Gras, Grassia, Gray, Gray, Greco, Green, Green, Gretarsson, Griggs,
  Grignani, Grimaldi, Grimm, Grote, Grunewald, Gruning, Guidi, Guimaraes,
  Guix{\'{e}}, Gulati, Guo, Gupta, Gupta, Gupta, Gustafson, Gustafson, Haegel,
  Halim, Hall, Hamilton, Hammond, Haney, Hanke, Hanks, Hanna, Hannam,
  Hannuksela, Hansen, Hanson, Harder, Hardwick, Haris, Harms, Harry, Harry,
  Hasskew, Haster, Haughian, Hayes, Healy, Heidmann, Heintze, Heinze, Heitmann,
  Hellman, Hello, Hemming, Hendry, Heng, Hennes, Hennig, Heurs, Hild, Hinderer,
  Hoback, Hochheim, Hofgard, Hofman, Holgado, Holland, Holt, Holz, Hopkins,
  Horst, Hough, Howell, Hoy, Huang, Hübner, Huerta, Huet, Hughey, Hui, Husa,
  Huttner, Huxford, Huynh-Dinh, Idzkowski, Iess, Inchauspe, Ingram, Intini,
  Isac, Isi, Iyer, Jacqmin, Jadhav, Jadhav, James, Jani, Janthalur, Jaranowski,
  Jariwala, Jaume, Jenkins, Jiang, Johns, Johnson-McDaniel, Jones, Jones,
  Jones, Jones, Jones, Jonker, Ju, Junker, Kalaghatgi, Kalogera, Kamai,
  Kandhasamy, Kang, Kanner, Kapadia, Karki, Kashyap, Kasprzack, Kastaun,
  Katsanevas, Katsavounidis, Katzman, Kaufer, Kawabe, K{\'{e}}f{\'{e}}lian,
  Keitel, Keivani, Kennedy, Key, Khadka, Khalili, Khan, Khan, Khan, Khazanov,
  Khetan, Khursheed, Kijbunchoo, Kim, Kim, Kim, Kim, Kim, Kim, Kim, Kimball,
  King, Kinley-Hanlon, Kirchhoff, Kissel, Kleybolte, Klimenko, Knowles,
  Knyazev, Koch, Koehlenbeck, Koekoek, Koley, Kondrashov, Kontos, Koper,
  Korobko, Korth, Kovalam, Kozak, Kringel, Krishnendu, Kr{\'{o}}lak, Krupinski,
  Kuehn, Kumar, Kumar, Kumar, Kumar, Kumar, Kuo, Kutynia, Lackey, Laghi,
  Lalande, Lam, Lamberts, Landry, Landry, Lane, Lang, Lange, Lantz, Lanza,
  Rosa, Lartaux-Vollard, Lasky, Laxen, Lazzarini, Lazzaro, Leaci, Leavey,
  Lecoeuche, Lee, Lee, Lee, Lee, Lee, Lehmann, Leroy, Letendre, Levin, Li, Li,
  li, Li, Li, Linde, Linker, Linley, Littenberg, Liu, Liu, Llorens-Monteagudo,
  Lo, Lockwood, London, Longo, Lorenzini, Loriette, Lormand, Losurdo, Lough,
  Lousto, Lovelace, Lück, Lumaca, Lundgren, Ma, Macas, Macfoy, MacInnis,
  Macleod, MacMillan, Macquet, Hernandez, Maga{\~{n}}a-Sandoval, Magee,
  Majorana, Maksimovic, Malik, Man, Mandic, Mangano, Mansell, Manske,
  Mantovani, Mapelli, Marchesoni, Marion, M{\'{a}}rka, M{\'{a}}rka, Markakis,
  Markosyan, Markowitz, Maros, Marquina, Marsat, Martelli, Martin, Martin,
  Martinez, Martynov, Masalehdan, Mason, Massera, Masserot, Massinger,
  Masso-Reid, Mastrogiovanni, Matas, Matichard, Mavalvala, Maynard, McCann,
  McCarthy, McClelland, McCormick, McCuller, McGuire, McIsaac, McIver, McManus,
  McRae, McWilliams, Meacher, Meadors, Mehmet, Mehta, Villa, Melatos, Mendell,
  Mercer, Mereni, Merfeld, Merilh, Merritt, Merzougui, Meshkov, Messenger,
  Messick, Metzdorff, Meyers, Meylahn, Mhaske, Miani, Miao, Michaloliakos,
  Michel, Middleton, Milano, Miller, Millhouse, Mills, Milotti, Milovich-Goff,
  Minazzoli, Minenkov, Mishkin, Mishra, Mistry, Mitra, Mitrofanov,
  Mitselmakher, Mittleman, Mo, Mogushi, Mohapatra, Mohite, Molina-Ruiz, Mondin,
  Montani, Moore, Moraru, Morawski, Moreno, Morisaki, Mours, Mow-Lowry, Mozzon,
  Muciaccia, Mukherjee, Mukherjee, Mukherjee, Mukherjee, Mukund, Mullavey,
  Munch, Mu{\~{n}}iz, Murray, Nagar, Nardecchia, Naticchioni, Nayak, Neil,
  Neilson, Nelemans, Nelson, Nery, Neunzert, Ng, Ng, Nguyen, Nguyen, Nichols,
  Nichols, Nissanke, Nocera, Noh, North, Nothard, Nuttall, Oberling, O'Brien,
  Oganesyan, Ogin, Oh, Oh, Ohme, Ohta, Okada, Oliver, Olivetto, Oppermann,
  Oram, O'Reilly, Ormiston, Ortega, O'Shaughnessy, Ossokine, Osthelder,
  Ottaway, Overmier, Owen, Pace, Pagano, Page, Pagliaroli, Pai, Pai, Palamos,
  Palashov, Palomba, Pan, Panda, Pang, Pankow, Pannarale, Pant, Paoletti,
  Paoli, Parida, Parker, Pascucci, Pasqualetti, Passaquieti, Passuello,
  Patricelli, Payne, Pearlstone, Pechsiri, Pedersen, Pedraza, Pele, Penn,
  Perego, Perez, P{\'{e}}rigois, Perreca, Perri{\`{e}}s, Petermann, Pfeiffer,
  Phelps, Phukon, Piccinni, Pichot, Piendibene, Piergiovanni, Pierro, Pillant,
  Pinard, Pinto, Piotrzkowski, Pirello, Pitkin, Plastino, Poggiani, Pong,
  Ponrathnam, Popolizio, Porter, Powell, Prajapati, Prasai, Prasanna, Pratten,
  Prestegard, Principe, Prodi, Prokhorov, Punturo, Puppo, Pürrer, Qi,
  Quetschke, Quinonez, Raab, Raaijmakers, Radkins, Radulesco, Raffai, Rafferty,
  Raja, Rajan, Rajbhandari, Rakhmanov, Ramirez, Ramos-Buades, Rana, Rao,
  Rapagnani, Raymond, Razzano, Read, Regimbau, Rei, Reid, Reitze, Rettegno,
  Ricci, Richardson, Richardson, Ricker, Riemenschneider, Riles, Rizzo,
  Robertson, Robinet, Rocchi, Rodriguez-Soto, Rolland, Rollins, Roma,
  Romanelli, Romano, Romel, Romero-Shaw, Romie, Rose, Rose, Rose,
  Rosi{\'{n}}ska, Rosofsky, Ross, Rowan, Rowlinson, Roy, Roy, Roy, Ruggi,
  Rutins, Ryan, Sachdev, Sadecki, Sakellariadou, Salafia, Salconi, Saleem,
  Salemi, Samajdar, Sanchez, Sanchez, Sanchis-Gual, Sanders, Santiago, Santos,
  Sarin, Sassolas, Sathyaprakash, Sauter, Savage, Savant, Sawant, Sayah,
  Schaetzl, Schale, Scheel, Scheuer, Schmidt, Schnabel, Schofield, Schönbeck,
  Schreiber, Schulte, Schutz, Schwarm, Schwartz, Scott, Scott, Seidel, Sellers,
  Sengupta, Sennett, Sentenac, Sequino, Sergeev, Setyawati, Shaddock, Shaffer,
  Shahriar, Sharma, Sharma, Shawhan, Shen, Shikauchi, Shink, Shoemaker,
  Shoemaker, Shukla, ShyamSundar, Siellez, Sieniawska, Sigg, Singer, Singh,
  Singh, Singha, Singhal, Sintes, Sipala, Skliris, Slagmolen, Slaven-Blair,
  Smetana, Smith, Smith, Somala, Son, Soni, Sorazu, Sordini, Sorrentino,
  Souradeep, Sowell, Spencer, Spera, Srivastava, Srivastava, Staats, Stachie,
  Standke, Steer, Steinhoff, Steinke, Steinlechner, Steinlechner, Steinmeyer,
  Stevenson, Stocks, Stops, Stover, Strain, Stratta, Strunk, Sturani, Stuver,
  Sudhagar, Sudhir, Summerscales, Sun, Sunil, Sur, Suresh, Sutton, Swinkels,
  Szczepa{\'{n}}czyk, Tacca, Tait, Talbot, Tanasijczuk, Tanner, Tao,
  T{\'{a}}pai, Tapia, Martin, Tasson, Taylor, Tenorio, Terkowski,
  Thirugnanasambandam, Thomas, Thomas, Thompson, Thondapu, Thorne, Thrane,
  Tinsman, Saravanan, Tiwari, Tiwari, Tiwari, Toland, Tonelli, Tornasi,
  Torres-Forn{\'{e}}, Torrie, e~Melo, Töyrä, Trail, Travasso, Traylor,
  Tringali, Tripathee, Trovato, Trudeau, Tsang, Tse, Tso, Tsukada, Tsuna,
  Tsutsui, Turconi, Ubhi, Ueno, Ugolini, Unnikrishnan, Urban, Usman, Utina,
  Vahlbruch, Vajente, Valdes, Valentini, van Bakel, van Beuzekom, van~den
  Brand, Broeck, Vander-Hyde, van~der Schaaf, Heijningen, van Veggel, Vardaro,
  Varma, Vass, Vas{\'{u}}th, Vecchio, Vedovato, Veitch, Veitch, Venkateswara,
  Venugopalan, Verkindt, Veske, Vetrano, Vicer{\'{e}}, Viets, Vinciguerra,
  Vine, Vinet, Vitale, Vivanco, Vo, Vocca, Vorvick, Vyatchanin, Wade, Wade,
  Wade, Walet, Walker, Wallace, Wallace, Walsh, Wang, Wang, Wang, Ward, Warden,
  Warner, Was, Watchi, Weaver, Wei, Weinert, Weinstein, Weiss, Wellmann, Wen,
  We{\ss}els, Westhouse, Wette, Whelan, Whiting, Whittle, Wilken, Williams,
  Willis, Willke, Winkler, Wipf, Wittel, Woan, Woehler, Wofford, Wong, Wright,
  Wu, Wysocki, Xiao, Yamamoto, Yang, Yang, Yang, Yap, Yazback, Yeeles, Yu, Yu,
  Yuen, Zadro{\.{z}}ny, Zadro{\.{z}}ny, Zanolin, Zelenova, Zendri, Zevin,
  Zhang, Zhang, Zhang, Zhao, Zhao, Zhou, Zhou, Zhu, Zimmerman, Zucker, \&
  Zweizig}]{Abbott2020APJ}
Abbott, R., Abbott, T.~D., Abraham, S., {et~al.} 2020, ApJL, 896, L44,
  \dodoi{10.3847/2041-8213/ab960f}

\bibitem[{Akmal {et~al.}(1998)Akmal, Pandharipande, \&
  Ravenhall}]{Akmal1998_PRC58-1804}
Akmal, A., Pandharipande, V.~R., \& Ravenhall, D.~G. 1998, PhRvC, 58, 1804,
  \dodoi{10.1103/PhysRevC.58.1804}

\bibitem[{Alonso \& Sammarruca(2003)}]{Alonso2003}
Alonso, D., \& Sammarruca, F. 2003, PhRvC, 67, 054301,
  \dodoi{10.1103/PhysRevC.67.054301}

\bibitem[{Anastasio {et~al.}(1981)Anastasio, Celenza, \&
  Shakin}]{Anastasio1981_PRC23-2273}
Anastasio, M.~R., Celenza, L.~S., \& Shakin, C.~M. 1981, PhRvC, 23, 2273,
  \dodoi{10.1103/PhysRevC.23.2273}

\bibitem[{Annala {et~al.}(2018)Annala, Gorda, Kurkela, \&
  Vuorinen}]{Annala2018}
Annala, E., Gorda, T., Kurkela, A., \& Vuorinen, A. 2018, PhRvL, 120, 172703,
  \dodoi{10.1103/PhysRevLett.120.172703}

\bibitem[{Antoniadis {et~al.}(2013)}]{antoniadis2013}
Antoniadis, J., {et~al.} 2013, Sci, 340, 1233232,
  \dodoi{10.1126/science.1233232}

\bibitem[{Baran {et~al.}(2005)Baran, Colonna, Greco, \& {Di Toro}}]{BARAN2005}
Baran, V., Colonna, M., Greco, V., \& {Di Toro}, M. 2005, PhR, 410, 335,
  \dodoi{https://doi.org/10.1016/j.physrep.2004.12.004}

\bibitem[{Baym {et~al.}(1971{\natexlab{a}})Baym, Bethe, \& Pethick}]{Baym19712}
Baym, G., Bethe, H.~A., \& Pethick, C.~J. 1971{\natexlab{a}}, NuPhA, 175, 225 ,
  \dodoi{https://doi.org/10.1016/0375-9474(71)90281-8}

\bibitem[{Baym {et~al.}(1971{\natexlab{b}})Baym, Pethick, \&
  Sutherland}]{Baym1971}
Baym, G., Pethick, C., \& Sutherland, P. 1971{\natexlab{b}}, ApJ, 170, 299,
  \dodoi{10.1086/151216}

\bibitem[{Bethe(1971)}]{Bethe1971}
Bethe, H.~A. 1971, Annu. Rev. Nucl. Sci., 21, 93,
  \dodoi{10.1146/annurev.ns.21.120171.000521}

\bibitem[{Boguta \& Bodmer(1977)}]{Boguta1977_NPA292-413}
Boguta, J., \& Bodmer, A. 1977, NuPhA, 292, 413,
  \dodoi{https://doi.org/10.1016/0375-9474(77)90626-1}

\bibitem[{Bombaci {et~al.}(2021)Bombaci, Drago, Logoteta, Pagliara, \&
  Vida\~na}]{BombaciPRL2021}
Bombaci, I., Drago, A., Logoteta, D., Pagliara, G., \& Vida\~na, I. 2021,
  PhRvL, 126, 162702, \dodoi{10.1103/PhysRevLett.126.162702}

\bibitem[{Bombaci \& Lombardo(1991)}]{Bombaci1991}
Bombaci, I., \& Lombardo, U. 1991, PhRvC, 44, 1892,
  \dodoi{10.1103/PhysRevC.44.1892}

\bibitem[{Brockmann \& Machleidt(1990)}]{Brockmann1990}
Brockmann, R., \& Machleidt, R. 1990, PhRvC, 42, 1965,
  \dodoi{10.1103/PhysRevC.42.1965}

\bibitem[{Brown {et~al.}(2018)Brown, Cumming, Fattoyev, Horowitz, Page, \&
  Reddy}]{BrownPRL2018}
Brown, E.~F., Cumming, A., Fattoyev, F.~J., {et~al.} 2018, PhRvL, 120, 182701,
  \dodoi{10.1103/PhysRevLett.120.182701}

\bibitem[{Burgio {et~al.}(2021)Burgio, Schulze, Vidaña, \&
  Wei}]{BURGIO2021PPNP}
Burgio, G., Schulze, H.-J., Vidaña, I., \& Wei, J.-B. 2021, PrPNP, 120,
  103879, \dodoi{https://doi.org/10.1016/j.ppnp.2021.103879}

\bibitem[{Centelles {et~al.}(2009)Centelles, Roca-Maza, Vi\~nas, \&
  Warda}]{Centelles2009}
Centelles, M., Roca-Maza, X., Vi\~nas, X., \& Warda, M. 2009, PhRvL, 102,
  122502, \dodoi{10.1103/PhysRevLett.102.122502}

\bibitem[{Chen {et~al.}(2005)Chen, Ko, \& Li}]{ChenKoLi2005}
Chen, L.-W., Ko, C.~M., \& Li, B.-A. 2005, PhRvC, 72, 064309,
  \dodoi{10.1103/PhysRevC.72.064309}

\bibitem[{Coester {et~al.}(1970)Coester, Cohen, Day, \& Vincent}]{Coester1970}
Coester, F., Cohen, S., Day, B., \& Vincent, C.~M. 1970, PhRvC, 1, 769,
  \dodoi{10.1103/PhysRevC.1.769}

\bibitem[{Cromartie {et~al.}(2020)}]{cromartie2020}
Cromartie, H.~T., {et~al.} 2020, NatAs, 4, 72,
  \dodoi{10.1038/s41550-019-0880-2}

\bibitem[{Damour \& Nagar(2009)}]{Damour2009}
Damour, T., \& Nagar, A. 2009, PhRvD, 80, 084035,
  \dodoi{10.1103/PhysRevD.80.084035}

\bibitem[{Damour {et~al.}(1992)Damour, Soffel, \& Xu}]{Damour1992}
Damour, T., Soffel, M., \& Xu, C. 1992, PhRvD, 45, 1017,
  \dodoi{10.1103/PhysRevD.45.1017}

\bibitem[{de~Jong \& Lenske(1998)}]{Jong1998}
de~Jong, F., \& Lenske, H. 1998, PhRvC, 57, 3099,
  \dodoi{10.1103/PhysRevC.57.3099}

\bibitem[{Decharg\'e \& Gogny(1980)}]{Decharge1980_PRC21-1568}
Decharg\'e, J., \& Gogny, D. 1980, PhRvC, 21, 1568,
  \dodoi{10.1103/PhysRevC.21.1568}

\bibitem[{Demorest {et~al.}(2010)Demorest, Pennucci, Ransom, Roberts, \&
  Hessels}]{demorest2010}
Demorest, P.~B., Pennucci, T., Ransom, S.~M., Roberts, M. S.~E., \& Hessels, J.
  W.~T. 2010, Natur, 467, 1081, \dodoi{10.1038/nature09466}

\bibitem[{Dong {et~al.}(2012)Dong, Zuo, Gu, \& Lombardo}]{DongPRC2012}
Dong, J., Zuo, W., Gu, J., \& Lombardo, U. 2012, PhRvC, 85, 034308,
  \dodoi{10.1103/PhysRevC.85.034308}

\bibitem[{Dutra {et~al.}(2012)Dutra, Louren\ifmmode~\mbox{\c{c}}\else
  \c{c}\fi{}o, S\'a~Martins, Delfino, Stone, \& Stevenson}]{Dutra2012}
Dutra, M., Louren\ifmmode~\mbox{\c{c}}\else \c{c}\fi{}o, O., S\'a~Martins,
  J.~S., {et~al.} 2012, PhRvC, 85, 035201, \dodoi{10.1103/PhysRevC.85.035201}

\bibitem[{Dutra {et~al.}(2014)Dutra, Louren\ifmmode~\mbox{\c{c}}\else
  \c{c}\fi{}o, Avancini, Carlson, Delfino, Menezes, Provid\^encia, Typel, \&
  Stone}]{Dutra2014}
Dutra, M., Louren\ifmmode~\mbox{\c{c}}\else \c{c}\fi{}o, O., Avancini, S.~S.,
  {et~al.} 2014, PhRvC, 90, 055203, \dodoi{10.1103/PhysRevC.90.055203}

\bibitem[{Engvik {et~al.}(1994)Engvik, Hjorth-Jensen, Osnes, Bao, \&
  \O{}stgaard}]{Engvik1994}
Engvik, L., Hjorth-Jensen, M., Osnes, E., Bao, G., \& \O{}stgaard, E. 1994,
  PhRvL, 73, 2650, \dodoi{10.1103/PhysRevLett.73.2650}

\bibitem[{Essick {et~al.}(2021)Essick, Tews, Landry, \&
  Schwenk}]{Essick2021PRL}
Essick, R., Tews, I., Landry, P., \& Schwenk, A. 2021, PhRvL, 127, 192701,
  \dodoi{10.1103/PhysRevLett.127.192701}

\bibitem[{Fattoyev {et~al.}(2020)Fattoyev, Horowitz, Piekarewicz, \&
  Reed}]{FattoyevPRC2020}
Fattoyev, F.~J., Horowitz, C.~J., Piekarewicz, J., \& Reed, B. 2020, PhRvC,
  102, 065805, \dodoi{10.1103/PhysRevC.102.065805}

\bibitem[{Fattoyev {et~al.}(2018)Fattoyev, Piekarewicz, \&
  Horowitz}]{Fattoyev2018PRL}
Fattoyev, F.~J., Piekarewicz, J., \& Horowitz, C.~J. 2018, PhRvL, 120, 172702,
  \dodoi{10.1103/PhysRevLett.120.172702}

\bibitem[{Flanagan \& Hinderer(2008)}]{Flanagan2008}
Flanagan, E.~E., \& Hinderer, T. 2008, PhRvD, 77, 021502(R),
  \dodoi{10.1103/PhysRevD.77.021502}

\bibitem[{Fonseca {et~al.}(2016)}]{Fonseca2016APJ}
Fonseca, E., {et~al.} 2016, ApJ, 832, 167, \dodoi{10.3847/0004-637X/832/2/167}

\bibitem[{Gandolfi {et~al.}(2012)Gandolfi, Carlson, \& Reddy}]{Gandolfi2012PRC}
Gandolfi, S., Carlson, J., \& Reddy, S. 2012, PhRvC, 85, 032801,
  \dodoi{10.1103/PhysRevC.85.032801}

\bibitem[{Garg \& Col{\`o}(2018)}]{Garg2018}
Garg, U., \& Col{\`o}, G. 2018, PrPNP, 101, 55,
  \dodoi{https://doi.org/10.1016/j.ppnp.2018.03.001}

\bibitem[{Garg {et~al.}(2007)Garg, Li, Okumura, Akimune, Fujiwara, Harakeh,
  Hashimoto, Itoh, Iwao, Kawabata, Kawase, Liu, Marks, Murakami, Nakanishi,
  Nayak, {Madhusudhana Rao}, Sakaguchi, Terashima, Uchida, Yasuda, Yosoi, \&
  Zenihiro}]{Garg2007}
Garg, U., Li, T., Okumura, S., {et~al.} 2007, NuPhA, 788, 36,
  \dodoi{https://doi.org/10.1016/j.nuclphysa.2007.01.046}

\bibitem[{Gross-Boelting {et~al.}(1999)Gross-Boelting, Fuchs, \&
  Faessler}]{GROSSBOELTING1999}
Gross-Boelting, T., Fuchs, C., \& Faessler, A. 1999, NuPhA, 648, 105,
  \dodoi{https://doi.org/10.1016/S0375-9474(99)00022-6}

\bibitem[{Hinderer(2008)}]{Hinderer2008}
Hinderer, T. 2008, ApJ, 677, 1216.
\newblock \url{http://stacks.iop.org/0004-637X/677/i=2/a=1216}

\bibitem[{Hofmann {et~al.}(2001)Hofmann, Keil, \&
  Lenske}]{Hofmann2001_PRC64-034314}
Hofmann, F., Keil, C.~M., \& Lenske, H. 2001, PhRvC, 64, 034314,
  \dodoi{10.1103/PhysRevC.64.034314}

\bibitem[{Horowitz \& Serot(1984)}]{HOROWITZ1984}
Horowitz, C., \& Serot, B.~D. 1984, PhLB, 137, 287,
  \dodoi{https://doi.org/10.1016/0370-2693(84)91717-9}

\bibitem[{Horowitz \& Serot(1987)}]{HOROWITZ1987}
---. 1987, NuPhA, 464, 613,
  \dodoi{https://doi.org/10.1016/0375-9474(87)90370-8}

\bibitem[{Huang {et~al.}(2020)Huang, Hu, Zhang, \& Shen}]{HuangApJ2020}
Huang, K., Hu, J., Zhang, Y., \& Shen, H. 2020, ApJ, 904, 39,
  \dodoi{10.3847/1538-4357/abbb37}

\bibitem[{Jiang {et~al.}(2015)Jiang, Yang, Dong, \& Long}]{Jiang2015PRC}
Jiang, L.~J., Yang, S., Dong, J.~M., \& Long, W.~H. 2015, PhRvC, 91, 025802,
  \dodoi{10.1103/PhysRevC.91.025802}

\bibitem[{Katayama \& Saito(2013)}]{Katayama2013}
Katayama, T., \& Saito, K. 2013, PhRvC, 88, 035805,
  \dodoi{10.1103/PhysRevC.88.035805}

\bibitem[{Krastev \& Sammarruca(2006)}]{Krastev2006}
Krastev, P.~G., \& Sammarruca, F. 2006, PhRvC, 74, 025808,
  \dodoi{10.1103/PhysRevC.74.025808}

\bibitem[{Kubis \& Kutschera(1997)}]{Kubis1997_PLB399-191}
Kubis, S., \& Kutschera, M. 1997, PhLB, 399, 191,
  \dodoi{https://doi.org/10.1016/S0370-2693(97)00306-7}

\bibitem[{Kumar \& Landry(2019)}]{kumar2019}
Kumar, B., \& Landry, P. 2019, PhRvD, 99, 123026,
  \dodoi{10.1103/PhysRevD.99.123026}

\bibitem[{Lattimer {et~al.}(1991)Lattimer, Pethick, Prakash, \&
  Haensel}]{Lattimer1991PRL}
Lattimer, J.~M., Pethick, C.~J., Prakash, M., \& Haensel, P. 1991, PhRvL, 66,
  2701, \dodoi{10.1103/PhysRevLett.66.2701}

\bibitem[{Lattimer \& Prakash(2004)}]{Lattimer2004}
Lattimer, J.~M., \& Prakash, M. 2004, Sci, 304, 536,
  \dodoi{10.1126/science.1090720}

\bibitem[{Lattimer \& Prakash(2007)}]{LATTIMER2007PR}
---. 2007, PhR, 442, 109, \dodoi{https://doi.org/10.1016/j.physrep.2007.02.003}

\bibitem[{Li {et~al.}(2008)Li, Chen, \& Ko}]{LiChenKo2008}
Li, B.-A., Chen, L.-W., \& Ko, C.~M. 2008, PhR, 464, 113,
  \dodoi{https://doi.org/10.1016/j.physrep.2008.04.005}

\bibitem[{Li \& Magno(2020)}]{BaoAnLi2020PRC}
Li, B.-A., \& Magno, M. 2020, PhRvC, 102, 045807,
  \dodoi{10.1103/PhysRevC.102.045807}

\bibitem[{Liu {et~al.}(2002)Liu, Greco, Baran, Colonna, \&
  Di~Toro}]{LIU-B2002_PRC65-045201}
Liu, B., Greco, V., Baran, V., Colonna, M., \& Di~Toro, M. 2002, PhRvC, 65,
  045201, \dodoi{10.1103/PhysRevC.65.045201}

\bibitem[{Long {et~al.}(2006)Long, Giai, \& Meng}]{Long2006}
Long, W.-H., Giai, N.~V., \& Meng, J. 2006, PhLB, 640, 150 ,
  \dodoi{https://doi.org/10.1016/j.physletb.2006.07.064}

\bibitem[{Machleidt(1989)}]{Machleidt1989}
Machleidt, R. 1989, Adv. Nucl. Phys., 19, 189,
  \dodoi{10.1007/978-1-4613-9907-0_2}

\bibitem[{Machleidt {et~al.}(1987)Machleidt, Holinde, \&
  Elster}]{Machleidt1987}
Machleidt, R., Holinde, K., \& Elster, C. 1987, PhR, 149, 1,
  \dodoi{https://doi.org/10.1016/S0370-1573(87)80002-9}

\bibitem[{Maieron {et~al.}(2004)Maieron, Baldo, Burgio, \&
  Schulze}]{Maieron2004}
Maieron, C., Baldo, M., Burgio, G.~F., \& Schulze, H.-J. 2004, PhRvD, 70,
  043010, \dodoi{10.1103/PhysRevD.70.043010}

\bibitem[{Malik {et~al.}(2018)Malik, Alam, Fortin, Provid\^encia, Agrawal, Jha,
  Kumar, \& Patra}]{Malik2018}
Malik, T., Alam, N., Fortin, M., {et~al.} 2018, PhRvC, 98, 035804,
  \dodoi{10.1103/PhysRevC.98.035804}

\bibitem[{Meng(2016)}]{CDFT2016}
Meng, J. 2016, Relativistic Density Functional for Nuclear Structure (WORLD
  SCIENTIFIC), \dodoi{10.1142/9872}

\bibitem[{Meng {et~al.}(2006)Meng, Toki, Zhou, Zhang, Long, \&
  Geng}]{Meng2006_PPNP57-470}
Meng, J., Toki, H., Zhou, S., {et~al.} 2006, PrPNP, 57, 470,
  \dodoi{https://doi.org/10.1016/j.ppnp.2005.06.001}

\bibitem[{Miller {et~al.}(2019)Miller, Lamb, Dittmann, Bogdanov, Arzoumanian,
  Gendreau, Guillot, Harding, Ho, Lattimer, Ludlam, Mahmoodifar, Morsink, Ray,
  Strohmayer, Wood, Enoto, Foster, Okajima, Prigozhin, \& Soong}]{Miller2019}
Miller, M.~C., Lamb, F.~K., Dittmann, A.~J., {et~al.} 2019, ApJ, 887, L24,
  \dodoi{10.3847/2041-8213/ab50c5}

\bibitem[{Most {et~al.}(2018)Most, Weih, Rezzolla, \&
  Schaffner-Bielich}]{Most2018}
Most, E.~R., Weih, L.~R., Rezzolla, L., \& Schaffner-Bielich, J. 2018, PhRvL,
  120, 261103, \dodoi{10.1103/PhysRevLett.120.261103}

\bibitem[{Newton \& Crocombe(2021)}]{Newton2021PRC}
Newton, W.~G., \& Crocombe, G. 2021, PhRvC, 103, 064323,
  \dodoi{10.1103/PhysRevC.103.064323}

\bibitem[{Nuppenau {et~al.}(1989)Nuppenau, Lee, \& MacKellar}]{NUPPENAU1989}
Nuppenau, C., Lee, Y., \& MacKellar, A. 1989, NuPhA, 504, 839,
  \dodoi{https://doi.org/10.1016/0375-9474(89)90011-0}

\bibitem[{Oppenheimer \& Volkoff(1939)}]{Oppenheimer1939}
Oppenheimer, J.~R., \& Volkoff, G.~M. 1939, PhRe, 55, 374,
  \dodoi{10.1103/PhysRev.55.374}

\bibitem[{Pethick(1992)}]{Pethick1992RMP}
Pethick, C.~J. 1992, RvMP, 64, 1133, \dodoi{10.1103/RevModPhys.64.1133}

\bibitem[{Poschenrieder \& Weigel(1988)}]{Poschenrieder1988_PRC38-471}
Poschenrieder, P., \& Weigel, M.~K. 1988, PhRvC, 38, 471,
  \dodoi{10.1103/PhysRevC.38.471}

\bibitem[{Postnikov {et~al.}(2010)Postnikov, Prakash, \&
  Lattimer}]{Postnikov2010}
Postnikov, S., Prakash, M., \& Lattimer, J.~M. 2010, PhRvD, 82, 024016,
  \dodoi{10.1103/PhysRevD.82.024016}

\bibitem[{Prakash(1994)}]{Prakash1994PR}
Prakash, M. 1994, PhR, 242, 297,
  \dodoi{https://doi.org/10.1016/0370-1573(94)90165-1}

\bibitem[{{Prakash} {et~al.}(1992){Prakash}, {Prakash}, {Lattimer}, \&
  {Pethick}}]{Prakash1992APJ}
{Prakash}, M., {Prakash}, M., {Lattimer}, J.~M., \& {Pethick}, C.~J. 1992,
  \apjl, 390, L77, \dodoi{10.1086/186376}

\bibitem[{Pudliner {et~al.}(1995)Pudliner, Pandharipande, Carlson, \&
  Wiringa}]{Pudliner1995_PRL74-4396}
Pudliner, B.~S., Pandharipande, V.~R., Carlson, J., \& Wiringa, R.~B. 1995,
  PhRvL, 74, 4396, \dodoi{10.1103/PhysRevLett.74.4396}

\bibitem[{Raaijmakers {et~al.}(2019)Raaijmakers, Riley, Watts, Greif, Morsink,
  Hebeler, Schwenk, Hinderer, Nissanke, Guillot, Arzoumanian, Bogdanov,
  Chakrabarty, Gendreau, Ho, Lattimer, Ludlam, \& Wolff}]{Raaijmakers_2019}
Raaijmakers, G., Riley, T.~E., Watts, A.~L., {et~al.} 2019, ApJ, 887, L22,
  \dodoi{10.3847/2041-8213/ab451a}

\bibitem[{Reinhard(1989)}]{Reinhard1989_RPP52-439}
Reinhard, P.~G. 1989, Rep. Prog. Phys., 52, 439,
  \dodoi{10.1088/0034-4885/52/4/002}

\bibitem[{Rhoades \& Ruffini(1974)}]{Rhoades1974PRL}
Rhoades, C.~E., \& Ruffini, R. 1974, PhRvL, 32, 324,
  \dodoi{10.1103/PhysRevLett.32.324}

\bibitem[{Riley {et~al.}(2019)Riley, Watts, Bogdanov, Ray, Ludlam, Guillot,
  Arzoumanian, Baker, Bilous, Chakrabarty, Gendreau, Harding, Ho, Lattimer,
  Morsink, \& Strohmayer}]{Riley_2019}
Riley, T.~E., Watts, A.~L., Bogdanov, S., {et~al.} 2019, ApJ, 887, L21,
  \dodoi{10.3847/2041-8213/ab481c}

\bibitem[{Ring(1996)}]{Ring1996_PPNP37-193}
Ring, P. 1996, PrPNP, 37, 193,
  \dodoi{https://doi.org/10.1016/0146-6410(96)00054-3}

\bibitem[{Roca-Maza {et~al.}(2011)Roca-Maza, Vi\~nas, Centelles, Ring, \&
  Schuck}]{RocaMaza2011_PRC84-054309}
Roca-Maza, X., Vi\~nas, X., Centelles, M., Ring, P., \& Schuck, P. 2011, PhRvC,
  84, 054309, \dodoi{10.1103/PhysRevC.84.054309}

\bibitem[{Safarzadeh \& Loeb(2020)}]{Safarzadeh2020ApJ}
Safarzadeh, M., \& Loeb, A. 2020, ApJ, 899, L15,
  \dodoi{10.3847/2041-8213/aba9df}

\bibitem[{Sagawa {et~al.}(2007)Sagawa, Yoshida, Zeng, Gu, \&
  Zhang}]{Sagawa2007}
Sagawa, H., Yoshida, S., Zeng, G.-M., Gu, J.-Z., \& Zhang, X.-Z. 2007, PhRvC,
  76, 034327, \dodoi{10.1103/PhysRevC.76.034327}

\bibitem[{Schiller \& M\"uther(2001)}]{Schiller2001}
Schiller, E., \& M\"uther, H. 2001, EPJA, 11, 15, \dodoi{10.1007/s100500170092}

\bibitem[{Serot \& Walecka(1986)}]{SerotWalecka1986}
Serot, B.~D., \& Walecka, J.~D. 1986, Adv. Nucl. Phys., 16, 1

\bibitem[{Shen {et~al.}(2019)Shen, Liang, Long, Meng, \& Ring}]{Shen2019}
Shen, S., Liang, H., Long, W.~H., Meng, J., \& Ring, P. 2019, PrPNP, 109,
  103713, \dodoi{https://doi.org/10.1016/j.ppnp.2019.103713}

\bibitem[{Skyrme(1956)}]{Skyrme1956}
Skyrme, T. H.~R. 1956, Phil. Mag., 1, 1043, \dodoi{10.1080/14786435608238186}

\bibitem[{Sprung(1972)}]{Sprung1972}
Sprung, D. W.~L. 1972, Adv. Nucl. Phys., 5, 225,
  \dodoi{10.1007/978-1-4615-8231-1_2}

\bibitem[{Sun {et~al.}(2008)Sun, Long, Meng, \& Lombardo}]{Sun2008}
Sun, B.~Y., Long, W.~H., Meng, J., \& Lombardo, U. 2008, PhRvC, 78, 065805,
  \dodoi{10.1103/PhysRevC.78.065805}

\bibitem[{Tews {et~al.}(2018)Tews, Margueron, \& Reddy}]{Tews2018}
Tews, I., Margueron, J., \& Reddy, S. 2018, PhRvC, 98, 045804,
  \dodoi{10.1103/PhysRevC.98.045804}

\bibitem[{Tolman(1939)}]{Tolman1939}
Tolman, R.~C. 1939, PhRe, 55, 364, \dodoi{10.1103/PhysRev.55.364}

\bibitem[{Tong {et~al.}(2018)Tong, Ren, Ring, Shen, Wang, \& Meng}]{Tong2018}
Tong, H., Ren, X.-L., Ring, P., {et~al.} 2018, PhRvC, 98, 054302,
  \dodoi{10.1103/PhysRevC.98.054302}

\bibitem[{Tong {et~al.}(2020)Tong, Zhao, \& Meng}]{Tong2020PRC}
Tong, H., Zhao, P., \& Meng, J. 2020, PhRvC, 101, 035802,
  \dodoi{10.1103/PhysRevC.101.035802}

\bibitem[{Tsokaros {et~al.}(2020)Tsokaros, Ruiz, \& Shapiro}]{Tsokaros2020ApJ}
Tsokaros, A., Ruiz, M., \& Shapiro, S.~L. 2020, ApJ, 905, 48,
  \dodoi{10.3847/1538-4357/abc421}

\bibitem[{Ulrych \& M\"uther(1997)}]{Ulrych1997_PRC56-1788}
Ulrych, S., \& M\"uther, H. 1997, PhRvC, 56, 1788,
  \dodoi{10.1103/PhysRevC.56.1788}

\bibitem[{{van Dalen} {et~al.}(2004){van Dalen}, Fuchs, \&
  Faessler}]{VanDalen2004_NPA744-227}
{van Dalen}, E., Fuchs, C., \& Faessler, A. 2004, NuPhA, 744, 227,
  \dodoi{https://doi.org/10.1016/j.nuclphysa.2004.08.019}

\bibitem[{van Dalen {et~al.}(2005)van Dalen, Fuchs, \&
  Faessler}]{VanDalen2005_PRL95-022302}
van Dalen, E. N.~E., Fuchs, C., \& Faessler, A. 2005, PhRvL, 95, 022302,
  \dodoi{10.1103/PhysRevLett.95.022302}

\bibitem[{Vattis {et~al.}(2020)Vattis, Goldstein, \&
  Koushiappas}]{VattisPRD2020}
Vattis, K., Goldstein, I.~S., \& Koushiappas, S.~M. 2020, PhRvD, 102, 061301,
  \dodoi{10.1103/PhysRevD.102.061301}

\bibitem[{Vautherin \& Brink(1972)}]{Vautherin1972_PRC5-626}
Vautherin, D., \& Brink, D.~M. 1972, PhRvC, 5, 626,
  \dodoi{10.1103/PhysRevC.5.626}

\bibitem[{Vida\~na {et~al.}(2009)Vida\~na, Provid\^encia, Polls, \&
  Rios}]{Vidana2009PRC}
Vida\~na, I., Provid\^encia, C., Polls, A., \& Rios, A. 2009, PhRvC, 80,
  045806, \dodoi{10.1103/PhysRevC.80.045806}

\bibitem[{Wang {et~al.}(2020)Wang, Hu, Zhang, \& Shen}]{Wang2020}
Wang, C., Hu, J., Zhang, Y., \& Shen, H. 2020, ApJ, 897, 96,
  \dodoi{10.3847/1538-4357/ab994b}

\bibitem[{Wang {et~al.}(2022)Wang, Tong, Zhao, Wang, Ring, \&
  Meng}]{SiboWang2022arxiv}
Wang, S., Tong, H., Zhao, Q., {et~al.} 2022, arXiv:2203.05397.
\newblock \doarXiv{2203.05397}

\bibitem[{Wang {et~al.}(2021)Wang, Zhao, Ring, \&
  Meng}]{WANG-SB2021_PRC103-054319}
Wang, S., Zhao, Q., Ring, P., \& Meng, J. 2021, PhRvC, 103, 054319,
  \dodoi{10.1103/PhysRevC.103.054319}

\bibitem[{Wiringa {et~al.}(1988)Wiringa, Fiks, \& Fabrocini}]{Wiringa1988PRC}
Wiringa, R.~B., Fiks, V., \& Fabrocini, A. 1988, PhRvC, 38, 1010,
  \dodoi{10.1103/PhysRevC.38.1010}

\bibitem[{Wiringa {et~al.}(1995)Wiringa, Stoks, \&
  Schiavilla}]{Wiringa1995_PRC51-38}
Wiringa, R.~B., Stoks, V. G.~J., \& Schiavilla, R. 1995, PhRvC, 51, 38,
  \dodoi{10.1103/PhysRevC.51.38}

\bibitem[{Xie \& Li(2020)}]{WenJieXie2020APJ}
Xie, W.-J., \& Li, B.-A. 2020, ApJ, 899, 4, \dodoi{10.3847/1538-4357/aba271}

\bibitem[{Yang {et~al.}(2020)Yang, Gayathri, Bartos, Haiman, Safarzadeh, \&
  Tagawa}]{Yang2020ApJ}
Yang, Y., Gayathri, V., Bartos, I., {et~al.} 2020, ApJ, 901, L34,
  \dodoi{10.3847/2041-8213/abb940}

\bibitem[{Zhang {et~al.}(2018)Zhang, Li, \& Xu}]{NaiBoZhang2018APJ}
Zhang, N.-B., Li, B.-A., \& Xu, J. 2018, ApJ, 859, 90,
  \dodoi{10.3847/1538-4357/aac027}

\bibitem[{{Zhang, Nai-Bo} \& {Li, Bao-An}(2019)}]{Zhang2019EPJA}
{Zhang, Nai-Bo}, \& {Li, Bao-An}. 2019, EPJA, 55, 39,
  \dodoi{10.1140/epja/i2019-12700-0}

\bibitem[{Zhao {et~al.}(2010)Zhao, Li, Yao, \& Meng}]{ZhaoLiYaoEtAl2010}
Zhao, P.~W., Li, Z.~P., Yao, J.~M., \& Meng, J. 2010, PhRvC, 82, 054319,
  \dodoi{10.1103/PhysRevC.82.054319}

\end{thebibliography}
\bibliographystyle{aasjournal}

\end{CJK*}
\end{document}